\pdfoutput=1
\RequirePackage{ifpdf}

\documentclass[cits,11pt,a4paper]{article}
\usepackage{jinstpub}

\usepackage{graphicx}
\usepackage{natbib}

\usepackage{lineno}

\title{High Voltage Insulation and Gas Absorption of Polymers in High Pressure Argon and Xenon Gases}

\usepackage{microtype}
\usepackage{rotating}
\usepackage{booktabs}
\usepackage{multirow}
\usepackage{tabularx}
\usepackage{dcolumn}
\usepackage{bm}
\usepackage[version=3]{mhchem}

\newcolumntype{R}{>{\raggedleft\arraybackslash}X}
\newcolumntype{Y}{>{\centering\arraybackslash}X}

\makeatletter
\g@addto@macro\bfseries{\boldmath}
\makeatother

\collaboration{The NEXT Collaboration}
\author[3,a]{L.~Rogers,\note[a]{Corresponding author.}}
\author[3]{R.A.~Clark,}
\author[3]{B.J.P.~Jones,}
\author[3]{A.D.~McDonald,}
\author[3,b]{D.R.~Nygren,\note[b]{NEXT Co-spokesperson.}}
\author[3]{F.~Psihas,}
\author[10]{C.~Adams,}
\author[17]{V.~\'Alvarez,}
\author[6]{L.~Arazi,}
\author[4]{C.D.R~Azevedo,}
\author[2]{K.~Bailey,}
\author[19]{F.~Ballester,}
\author[17]{J.M.~Benlloch-Rodr\'{i}guez,}
\author[12]{F.I.G.M.~Borges,}
\author[17]{A.~Botas,}
\author[17]{S.~C\'arcel,}
\author[17]{J.V.~Carri\'on,}
\author[20]{S.~Cebri\'an,}
\author[12]{C.A.N.~Conde,}
\author[17]{J.~D\'iaz,}
\author[5]{M.~Diesburg,}
\author[12]{J.~Escada,}
\author[19]{R.~Esteve,}
\author[17]{R.~Felkai,}
\author[11]{A.F.M.~Fernandes,}
\author[11]{L.M.P.~Fernandes,}
\author[14,8,17]{P.~Ferrario,}
\author[4]{A.L.~Ferreira,}
\author[11]{E.D.C.~Freitas,}
\author[14]{J.~Generowicz,}
\author[7]{A.~Goldschmidt,}
\author[14,8,17,c]{J.J.~G\'omez-Cadenas,\note[c]{NEXT Co-spokesperson.}}
\author[18]{D.~Gonz\'alez-D\'iaz,}
\author[10]{R.~Guenette,}
\author[9]{R.M.~Guti\'errez,}
\author[2]{K.~Hafidi,}
\author[1]{J.~Hauptman,}
\author[11]{C.A.O.~Henriques,}
\author[9]{A.I.~Hernandez,}
\author[18]{J.A.~Hernando~Morata,}
\author[19]{V.~Herrero,}
\author[2]{S.~Johnston,}
\author[17]{M.~Kekic,}
\author[16]{L.~Labarga,}
\author[17]{A.~Laing,}
\author[5]{P.~Lebrun,}
\author[17]{N.~L\'opez-March,}
\author[9]{M.~Losada,}
\author[11]{R.D.P.~Mano,}
\author[10]{J.~Mart\'in-Albo,}
\author[17]{A.~Mart\'inez,}
\author[17,18]{G.~Mart\'inez-Lema,}
\author[3,14]{F.~Monrabal,}
\author[11]{C.M.B.~Monteiro,}
\author[19]{F.J.~Mora,}
\author[17]{J.~Mu\~noz Vidal,}
\author[17]{M.~Musti,}
\author[17]{M.~Nebot-Guinot,}
\author[17]{P.~Novella,}
\author[17]{B.~Palmeiro,}
\author[5]{A.~Para,}
\author[17,d]{J.~P\'erez,\note[d]{Now at Laboratorio Subterr\'aneo de Canfranc, Spain.}}
\author[17]{M.~Querol,}
\author[17]{J.~Renner,}
\author[2]{J.~Repond,}
\author[2]{S.~Riordan,}
\author[15]{L.~Ripoll,}
\author[17]{J.~Rodr\'iguez,}
\author[17]{C.~Romo-Luque,}
\author[12]{F.P.~Santos,}
\author[11]{J.M.F. dos~Santos,}
\author[17,6]{A.~Sim\'on,}
\author[13,e]{C.~Sofka,\note[e]{Now at University of Texas at Austin, USA.}}
\author[17]{M.~Sorel,}
\author[13]{T.~Stiegler,}
\author[19]{J.F.~Toledo,}
\author[17]{J.~Torrent,}
\author[4]{J.F.C.A.~Veloso,}
\author[13]{R.~Webb,}
\author[13,f]{J.T.~White,\note[f]{Deceased.}}
\author[17]{N.~Yahlali}
\emailAdd{leslie.rogers@mavs.uta.edu}
\affiliation[1]{
Department of Physics and Astronomy, Iowa State University, 12 Physics Hall, Ames, IA 50011-3160, USA}
\affiliation[2]{
Argonne National Laboratory, Argonne, IL 60439, USA}
\affiliation[3]{
Department of Physics, University of Texas at Arlington, Arlington, TX 76019, USA}
\affiliation[4]{
Institute of Nanostructures, Nanomodelling and Nanofabrication (i3N), Universidade de Aveiro, Campus de Santiago, Aveiro, 3810-193, Portugal}
\affiliation[5]{
Fermi National Accelerator Laboratory, Batavia, IL 60510, USA}
\affiliation[6]{
Nuclear Engineering Unit, Faculty of Engineering Sciences, Ben-Gurion University of the Negev, P.O.B. 653, Beer-Sheva, 8410501, Israel}
\affiliation[7]{
Lawrence Berkeley National Laboratory (LBNL), 1 Cyclotron Road, Berkeley, CA 94720, USA}
\affiliation[8]{
Ikerbasque, Basque Foundation for Science, Bilbao, E-48013, Spain}
\affiliation[9]{
Centro de Investigaci\'on en Ciencias B\'asicas y Aplicadas, Universidad Antonio Nari\~no, Sede Circunvalar, Carretera 3 Este No.\ 47 A-15, Bogot\'a, Colombia}
\affiliation[10]{
Department of Physics, Harvard University, Cambridge, MA 02138, USA}
\affiliation[11]{
LIBPhys, Physics Department, University of Coimbra, Rua Larga, Coimbra, 3004-516, Portugal}
\affiliation[12]{
LIP, Department of Physics, University of Coimbra, Coimbra, 3004-516, Portugal}
\affiliation[13]{
Department of Physics and Astronomy, Texas A\&M University, College Station, TX 77843-4242, USA}
\affiliation[14]{
Donostia International Physics Center (DIPC), Paseo Manuel Lardizabal, 4, Donostia-San Sebastian, E-20018, Spain}
\affiliation[15]{
Escola Polit\`ecnica Superior, Universitat de Girona, Av.~Montilivi, s/n, Girona, E-17071, Spain}
\affiliation[16]{
Departamento de F\'isica Te\'orica, Universidad Aut\'onoma de Madrid, Campus de Cantoblanco, Madrid, E-28049, Spain}
\affiliation[17]{
Instituto de F\'isica Corpuscular (IFIC), CSIC \& Universitat de Val\`encia, Calle Catedr\'atico Jos\'e Beltr\'an, 2, Paterna, E-46980, Spain}
\affiliation[18]{
Instituto Gallego de F\'isica de Altas Energ\'ias, Univ.\ de Santiago de Compostela, Campus sur, R\'ua Xos\'e Mar\'ia Su\'arez N\'u\~nez, s/n, Santiago de Compostela, E-15782, Spain}
\affiliation[19]{
Instituto de Instrumentaci\'on para Imagen Molecular (I3M), Centro Mixto CSIC - Universitat Polit\`ecnica de Val\`encia, Camino de Vera s/n, Valencia, E-46022, Spain}
\affiliation[20]{
Laboratorio de F\'isica Nuclear y Astropart\'iculas, Universidad de Zaragoza, Calle Pedro Cerbuna, 12, Zaragoza, E-50009, Spain}
\abstract{High pressure gas time projection chambers (HPGTPCs) are made with a variety of materials, many of which still await proper characterization in high pressure noble gas environments. As HPGTPCs increase in size toward ton-scale detectors, assemblies become larger and more complex, creating a need for detailed understanding of how structural supports and high voltage insulators behave. This includes identification of materials with predictable mechanical properties and without surface charge accumulation that may lead to field deformation or sparking. This paper explores the mechanical and electrical effects of high pressure gas environments on insulating polymers PTFE, HDPE, PEEK, POM and UHMW in argon and xenon, including studying gas absorption, swelling and high voltage insulation strength.}


\begin{document}
\maketitle

\section{High pressure gas time projection chambers}
High pressure gas time projection chambers (HPGTPC) are powerful detectors used in searches for neutrinoless double beta decay \cite{Martin-Albo:2015rhw,Chen:2016qcd,Luscher:1998sd} and neutrino oscillation measurements \cite{Martin-Albo:2016tfh,Hamilton:2015ows}. Advantages of HPGTPCs include precise energy and spatial resolution that enables event topology discrimination. This combination of energy measurements and event topology allows for particle identification resulting in strong background rejections \cite{Renner:2016trj,Ferrario:2015kta}. HPGTPCs use a drift region with a typical electric field strength of 200-500 V/cm to induce electron transport to the anode plane. Electrons reach the anode after a finite drift time, with this time serving as a proxy for z position. The 2D location of charge in the electroluminescent (EL) plane provides the x and y components. Taking these together provides a 3D reconstruction of the initial event.  This is the basis for the time projection chamber concept \cite{Fancher:1978es}.

%define HPGTPC topological ^^

Several technical variations of HPGTPCs exist. Due to its high price, xenon is generally only used when it has a very specific benefit, such as searching for lepton number violation. Xenon 136 is a candidate double beta decay isotope, because energy conservation allows for the double beta decay into Barium, while disallowing single beta decay into Cesium \cite{Avignone:2007fu}. 

\begin{figure}[t]
    \centering

    \includegraphics[width=.7\linewidth]{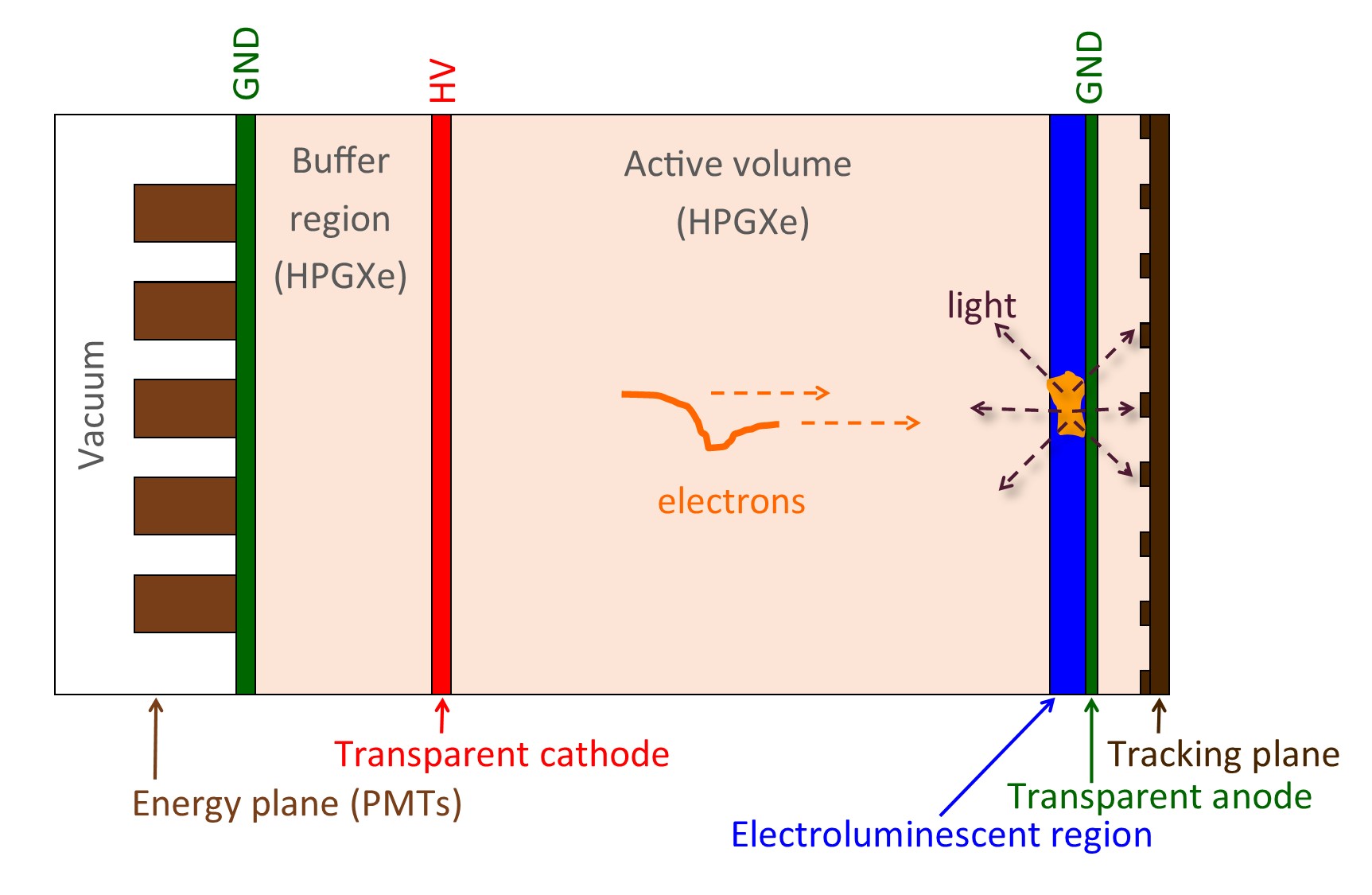}
    \caption{Schematic of an asymmetrical NEXT detector with PMTs on the left which collect light for an energy reading. A buffer region is used for stepping voltage up to the strength needed at the cathode and the drift region then steps voltage across to the EL region. Behind the EL region is an array of SiPMs to collect light for topological reconstruction.}    
    \label{fig:NEXT}
\end{figure}

The NEXT program \cite{Alvarez:2012flf,Granena:2009it,Gomez-Cadenas:2013lta,Alvarez:2012yxw,Alvarez:2012zsz} is a sequence of high pressure xenon gas TPCs. The existing detector designs use an asymmetrical configuration as shown in Fig.~\ref{fig:NEXT}, running at 10-15 bar. When ionization electrons reach the end of the drift region, they enter an EL region which has a high electric field that sufficiently accelerates charges to excite the noble gas atoms, but not ionize them. With appropriate tuning of this field strength, near-fluctationless gain can be achieved by collecting the copious photons that are produced as the excited atoms return to the ground state. Photomultiplier tubes (PMTs) and/or silicon photomultipliers (SiPMs) are used to collect the photons, following a wavelength shift by TPB \cite{Alvarez:2012ub} coating the outside walls of the active volume. The initial photons created from the scintillation (S1) are used as a trigger for the event, whereas the photons from the excitations in the EL region (S2) are collected for a precise energy reading and topological event characterization.

\begin{figure}[t]
    \centering
   \includegraphics[width=0.8\linewidth]{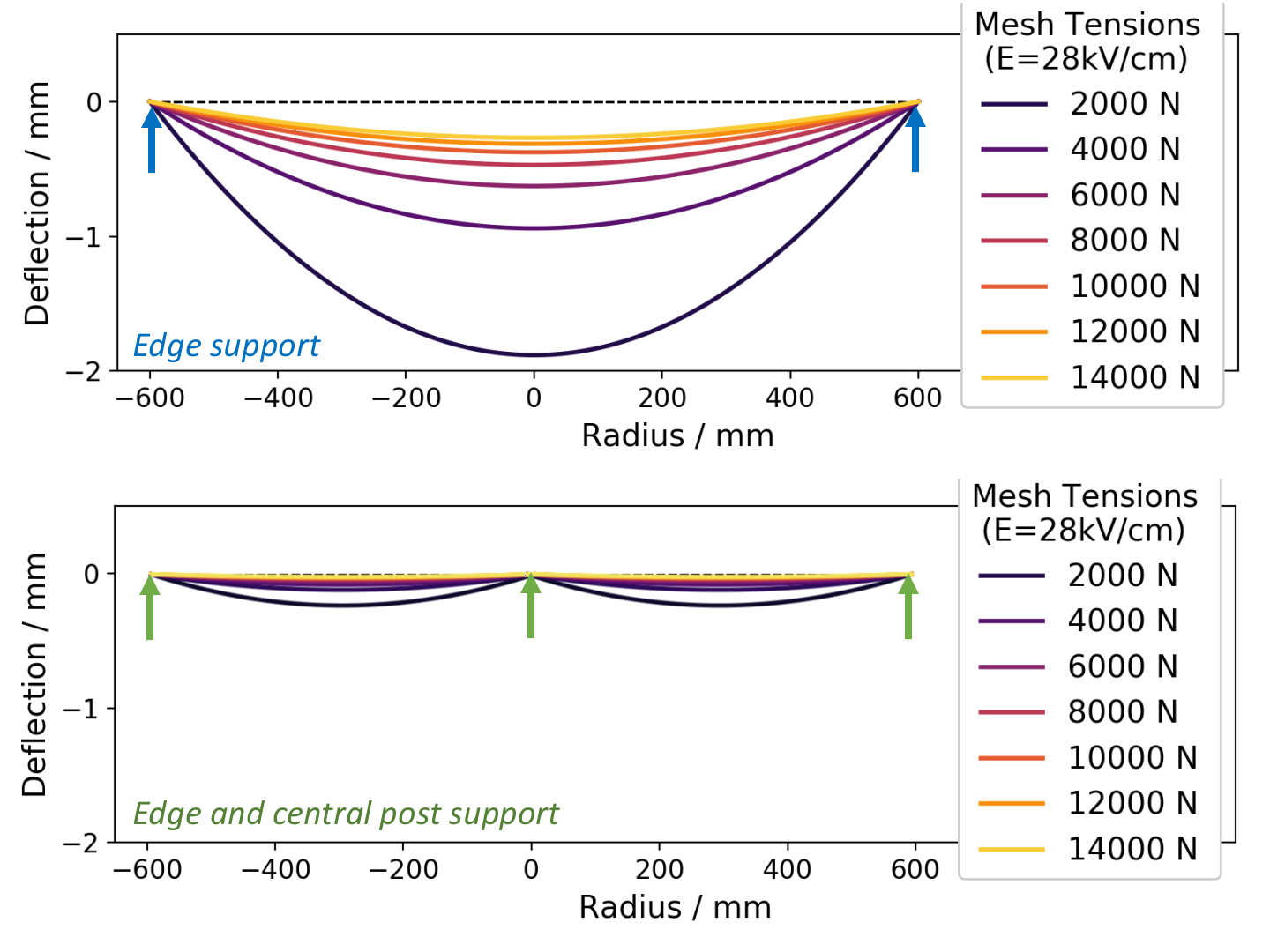}
    \caption{Simulated EL mesh deflection due to electrostatic forces with only edge support (top) or with edge support plus a single central support post (bottom). Dramatic deflection reductions are observed with only a small number of support posts, thus maintaining EL field and gain uniformity.}    
    \label{fig:MagPosts}
\end{figure}

A particular challenge in scaling detectors of this type for ton scale experiments lies in EL regions which require a very high and well controlled reduced electric field on the order of 2-4 kV cm$^{-1}$bar$^{-1}$ for argon and xenon \cite{monteiro2011determination}. This involves biasing large surfaces which must be transparent to 400 nm light, in order not to block the tracking or energy plane. The cathode side of the EL region also must be transparent to electrons, and so is commonly made with a thin wire mesh with high optical and electron transparency. At these large fields of around 30~kV/cm, the electrostatic force between the two electrodes may become sufficiently large as to cause significant deflection in a wire mesh at achievable tensions. Since an electric field is a function of voltage over distance between the two planes, deflection in the mesh causes the region to have a nonuniform field.  Calculated deflections at 28kV/cm for different mesh tensions in a NEXT-100 sized mesh are shown in Fig.~\ref{fig:MagPosts}, top.

To avoid such deflections, support materials may be introduced sporadically between the gate and anode Fig.~\ref{fig:MagPosts}, bottom. This material introduces non-conducting surfaces into the EL region, and there is a risk that this may encourage sparks between the electrodes.  This must be avoided, since as mesh size grows, the stored electrostatic energy increases as the square of the radius.  In the event of a spark, this energy is discharged through a single point, which for larger meshes may cause damage through localized heating.  These sparks may be nucleated by the accumulation of charges on insulating surfaces. The ideal material is thus dielectrically strong but somewhat resistive, in order to maintain a uniform axial field in the EL gap and avoid charge-up effects.  Such discharges may be quenched by other means, for example, by replacing the anode with a resistive material.  

As well as NEXT, other examples of HPGTPC experiments include; Gotthard \cite{Luscher:1998sd}, a pioneering predecessor for HPGXe neutrinoless double beta decay experiments. Gotthard was an asymmetrical xenon TPC with 4\% CH$_{4}$ at 5 bar that read charges out via a multi-wire proportional chamber rather than SiPMs or PMTs and used the same plane for both energy and tracking;  AXEL \cite{Pan:2016pbm}, which is similar to NEXT with 10 bar xenon, asymmetrical and with an EL region; and, PANDAX-III \cite{Chen:2016qcd}, a proposed 10 bar xenon TPC which is symmetrical with the cathode in the middle and drift towards both ends of the detector and charge readout via Micromegas, with TMA added as a quench gas.

When a large active volume is needed for viewing interactions, argon is a more cost effective medium.  A proposed high-pressure argon near detector for the DUNE experiment \cite{Martin-Albo:2016tfh} would be used to characterize the neutrino beam at short baseline in order to facilitate a precise neutrino oscillation search. This would be classified as an asymmetrical HPGTPC with a large drift region. 

A persistent challenge for noble TPC detectors is the capability to stably apply high voltages to produce constant drift fields over long distances.  The difficulty of creating this field naturally increases as experiments grow in size and the maximal high voltage increases.  Unfortunately, the importance of reaching the design field also increases, as  drift distances become longer and finite free electron lifetime due to attachment on impurities becomes a limiting factor.  Many major TPC experiments in both liquid and gas phase have failed to achieve their design field strengths, and so better understanding of insulating materials and their electrical breakdown properties is of much interest. 

EL TPCs must, in addition to the drift field, apply an EL amplification field.  While the EL field strength does not change between different detector scales, for larger detectors the maintenance of EL region planarity without deflection also becomes an electro-mechanical challenge.  The work in this paper is motivated by the need to satisfy the material requirements of EL support structures in NEXT-100 and NEXT-ton scale detectors, and may also inform HV design of drift regions where insulating materials are used. We investigate the behaviour of various insulating materials in high pressure argon and xenon gases, under large applied voltages at $\sim$ cm length scales. Previously uncharacterized swelling behaviour through xenon gas absorption is studied, and dielectric strength measurements transverse to the insulating surfaces are made.

This paper is organized as follows. Sec.~\ref{sec:Polymers} describes the materials under consideration and some previously used gases in noble TPC experiments.  Sec.~\ref{sec:Mechanical} reports on swelling through absorption of argon or xenon gases in high pressure environments; Sec.~\ref{sec:Electrical} reports on measurements of surface strength under applied transverse high voltage.  Finally, in Sec.~\ref{sec:Discussion} we discuss our findings and their implications for HPGTPC experiments, and specifically as support material for the NEXT-100 and NEXT-ton scale EL regions.

\section{Polymers as HV insulators in TPCs \label{sec:Polymers}}    
A variety of polymers are used in HPGTPCs as both structural components and as insulation from the high voltages required to produce high electric fields. These materials are required to have a low outgassing rate to avoid contaminating the noble gas as well as generally needing to satisfy strict radiopurity requirements. The ideal materials for HV applications are insulators with sufficient leakage current to avoid charge up effects that can distort electric fields or cause sparking.  For other applications, properties such as reflectivity and structural strength may also be design drivers.  This paper characterizes the behavior of some commonly used structural polymers that may find application in HPGTPCs, with a particular focus on candidates for EL support material in NEXT-100.  In this section we describe the materials under consideration and review some past use cases.

{\bf High Density Polyethylene (HDPE)} is a commonly used thermoplastic which has found past application as a structural field cage material in the NEXT-NEW \cite{March:2018vsx} experiment.  It has also been used as an insulating material in high voltage feedthroughs for use in liquid argon R\&D towards DUNE \cite{Cantini:2016tfx}, as cable insulation in the liquid xenon LZ experiment \cite{Mount:2017qzi} and as a shield material for XENON10 \cite{Aprile:2010bt}.

{\bf Polytetrafluoroethylene (PTFE)}, widely known under trade name Teflon, is a fluoropolymer which is often employed in xenon time projection chambers because of its properties as a strong diffuse reflector. It has been used for this purpose in the NEXT-NEW \cite{March:2018vsx}  HPGXeTPC and the LUX \cite{Akerib:2012ys} and EXO-200 \cite{Auger:2012gs} liquid xenon detectors.  Teflon has also been used in HV feedthrough applications in liquid xenon, forming a large part of the insulating material for the feedthroughs of the EXO-200 \cite{Auger:2012gs}. PTFE serves both structural and reflective purposes as the field cage frame in the xenon detectors \cite{Aprile:2010bt} and LZ \cite{Mount:2017qzi} experiment. The properties of PTFE surface breakdowns in liquid argon were recently studied in \cite{Lockwitz:2015qua}.

\begin{figure}[t]
\centering
\includegraphics[height=0.35\linewidth]{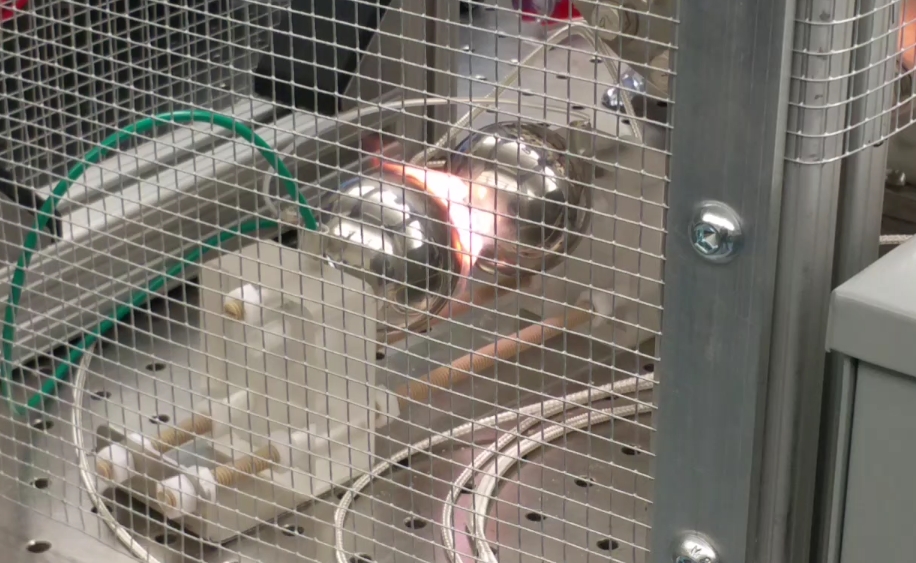} \includegraphics[height=0.35\linewidth]{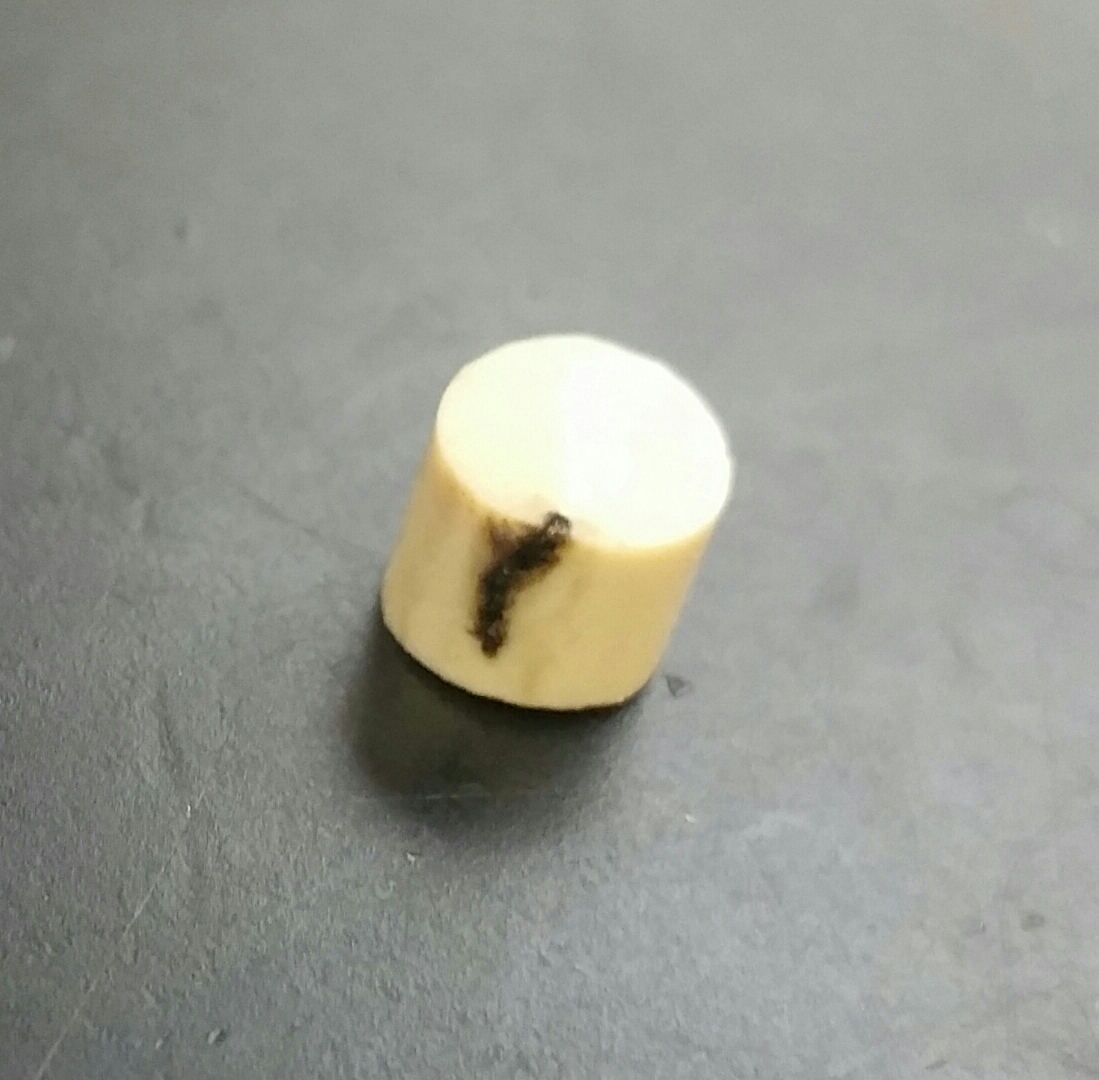}
\caption{Left: anti-static UHMW tested in air and failing through sparking and excessive heating at $\leq$10~kV; Right: picture of PEEK post that had a destructive spark leave a track along the side.}
\label{fig:destruction}
\end{figure}

{\bf Ultra-high molecular weight polyethylene (UHMW)} is a higher density, softer polyethylene material than HDPE, which has found wide use in high voltage feedthrough applications in liquid argon detectors including the ICARUS T600 \cite{Amerio:2004ze} and MicroBooNE \cite{Acciarri:2016smi} detectors. It is also used in structural field cage components of the LUX experiment \cite{Akerib:2012ys}.  The properties of UHMW surface breakdowns in liquid argon were recently studied in \cite{Lockwitz:2015qua}. In addition to virgin UHMW we also tested {\bf antistatic UHMW}, sometimes known by the trade name Tivar. This is formed by adding carbon powder to the UHMW resin to achieve surface resistivities of 10$^9\Omega$/sq.  It was anticipated that this surface resistivity would help avoid charge-up effects and sparking, although poor performance was obtained in preliminary tests.  In a preliminary scan of material dielectric strength in air, antistatic UHMW exhibited surface breakdowns at low voltages yielding an electric current far in excess of those expected based on surface resistivity. This led to destructive heating under applied voltages of less than 10~kV, disqualifying antistatic UHMW from further tests (Fig.~\ref{fig:destruction}, left).

{\bf Polyether ether ketone (PEEK)} is a high-performance engineering polymer commonly used for its robust mechanical properties and ultra high vacuum compatibility. It is extremely machinable and durable within a wide temperature range. PEEK has been used as a structural material in the WArP liquid argon dark matter search \cite{zani2014warp} and the PandaX \cite{cao2014pandax} experiment, as well as an insulator in the feed-throughs of the liquid XENON10  detector \cite{Aprile:2010bt}.   PEEK has become relatively disfavored as a field cage material for ultra low background experiments, since radioactivity of recently tested samples has been quantified at higher levels than other candidate materials \cite{Alvarez:2012as}.  Notably, PEEK was also the only material which failed in a destructive way during our high pressure gas HV strength tests, as will be described later (Fig.~\ref{fig:destruction}, right).

{\bf Polyoxymethylene (POM)}, often known as acetal or by trade names such as Delrin, is a machinable structural polymer.  We are not aware of previous uses in TPC experiments, although its surface electrical strength was recently characterized in liquid argon \cite{Lockwitz:2015qua}.  Promising performance there motivated us to explore its potential for use in HPGTPC experiments.

\begin{figure}[t]
    \centering
    \includegraphics[width=0.49\linewidth]{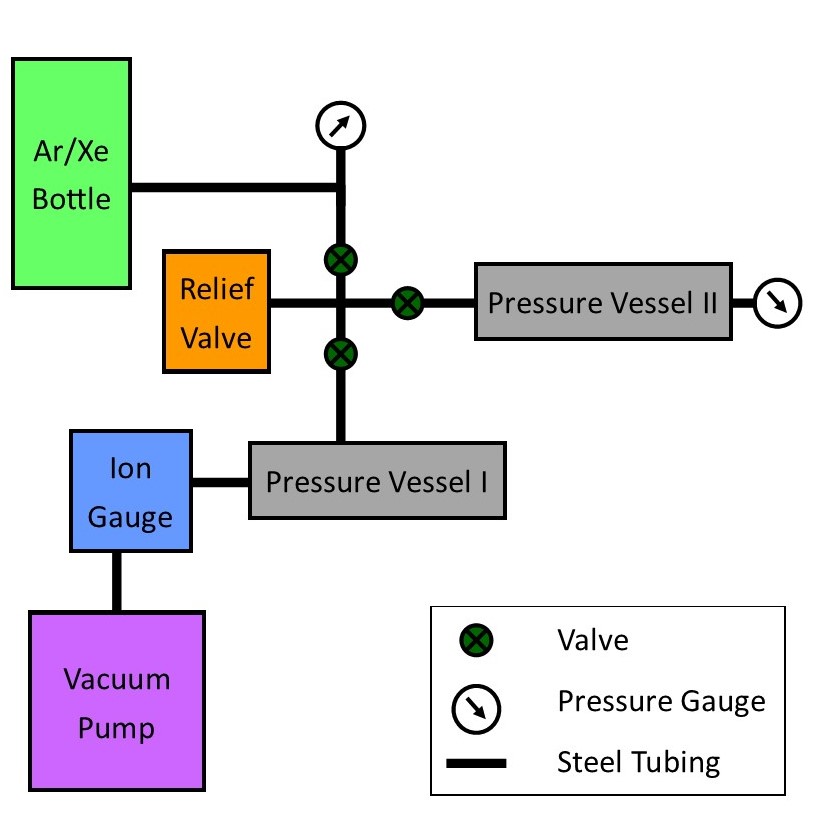}
    \includegraphics[angle=90,width=0.46\linewidth]{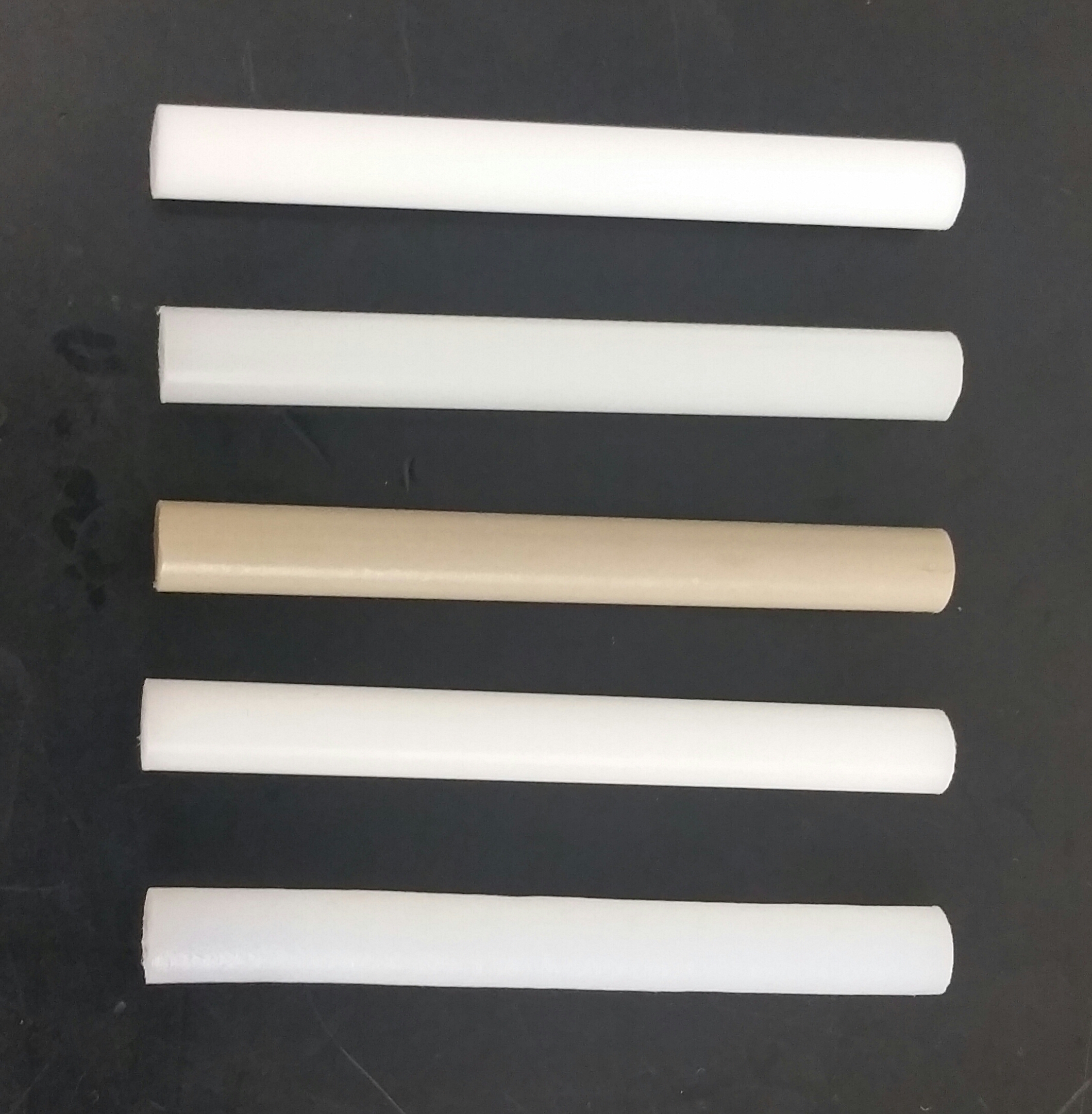}
    \caption{Left: Schematic of the xenon absorption test set up. Pressure Vessel I was where the evacuation tests were performed and Pressure Vessel II was where the rods were soaked in noble gas. Pressure Vessel II could be closed off with its valve and removed from the system, allowing multiple vessels to run at the same time.  Right: Typical posts used for these tests. From left to right the materials are PTFE, HDPE, PEEK, POM, and UHMW.  }    
    \label{fig:AbsorpSetup}
\end{figure}

\section{Tests of mechanical effects of gas absorption \label{sec:Mechanical}}
\subsection{Methodology}
To test the degree of swelling for polymers of interest, 6 cm length sections were cut from 6.4 mm diameter rods of these materials on a band saw and the edges deburred (Fig.~\ref{fig:AbsorpSetup}, right). The rods were cleaned for 15 minutes in ethyl-alcohol placed in a sonic bath. After taking the rods out and letting them dry, each was numbered with an identifying code. Each rod was measured in length 5 times with a pair of vernier calipers and 3 times in mass with a digital analytical balance scale. Each rod's measured masses and lengths were averaged and their standard deviation calculated yielding an average precision of 0.04 mm in length and 2 mg in mass.

The pieces were put in a small, 0.15 liter pressure cylinder, evacuated overnight, weighed, and measured again, with no perceptible change in either quantity. The rods were placed back inside the cylinder, evacuated to 4x\(10^{-5}\) Torr, and then pressurized with noble gas and left sealed until ready to re-measure. This setup is shown schematically in Fig.~\ref{fig:AbsorpSetup}, left. To test pressure dependence, rods were soaked under 5, 10, and 15 bar of xenon gas for 1 week each. Tests for time and noble gas dependence were performed by soaking rods in 10 bar of xenon and 10 bar of argon for 3.5, 7, 10, and 21 days.  Significant swelling was observed in some cases, as will be described below.

To establish if the observed swelling was permanent, posts that had soaked for a week in 15 bar of xenon were evacuated. Every few days the lengths and masses were re-measured and compared with their initial values. The system vacuum was monitored and observed to improve over the 14 days of the study as the rods out-gassed the residual xenon and lost mass and length.

Each set of runs had control posts that were kept in a sealed container in atmospheric air and measured at the same times as the test pieces. All measurements were reported as the new mass (or length) divided by the original, so that no change would be equivalent to 1.0.  This compensates for any bias in our weighing or measuring procedures. The controls were used to correct for measurement device variations by dividing by the average control ratio.

Up to 5 posts were measured for each material, and the length and mass changes are reported as mean and standard deviation over these posts.  The uncertainties on the ratios include a contribution from the control measurements, and so include some correlated uncertainty between points taken in the same run.

%\begin{equation}
%{\mu_0=\frac{\mu_{1}+\mu_{2}+...+\mu_{N}}{N}}
%\end{equation}

%\begin{equation}
%{\sigma_0=\sqrt{\frac{\sigma_{1}^{2}+\sigma_{2}^{2}+...+\sigma_{N}^{2}}{N}}}
%\end{equation}

%The ratios of the masses (or lengths) and the %corresponding errors were then found as
%\begin{equation}
%{\mu_R=\frac{\mu_{1}}{\mu_{0}}}
%\end{equation}

%\begin{equation}
%{\sigma_R=\sqrt{\frac{\sigma_{1}}{\mu_{1}}^{2}+\frac{\sigma_{0}}{\mu_{0}}^{2}}\left(\frac{\mu_{1}}{\mu_{0}}\right)}
%\end{equation}

\subsection{Results}

\begin{figure}[t]
\centering
\includegraphics[width=0.49\linewidth]{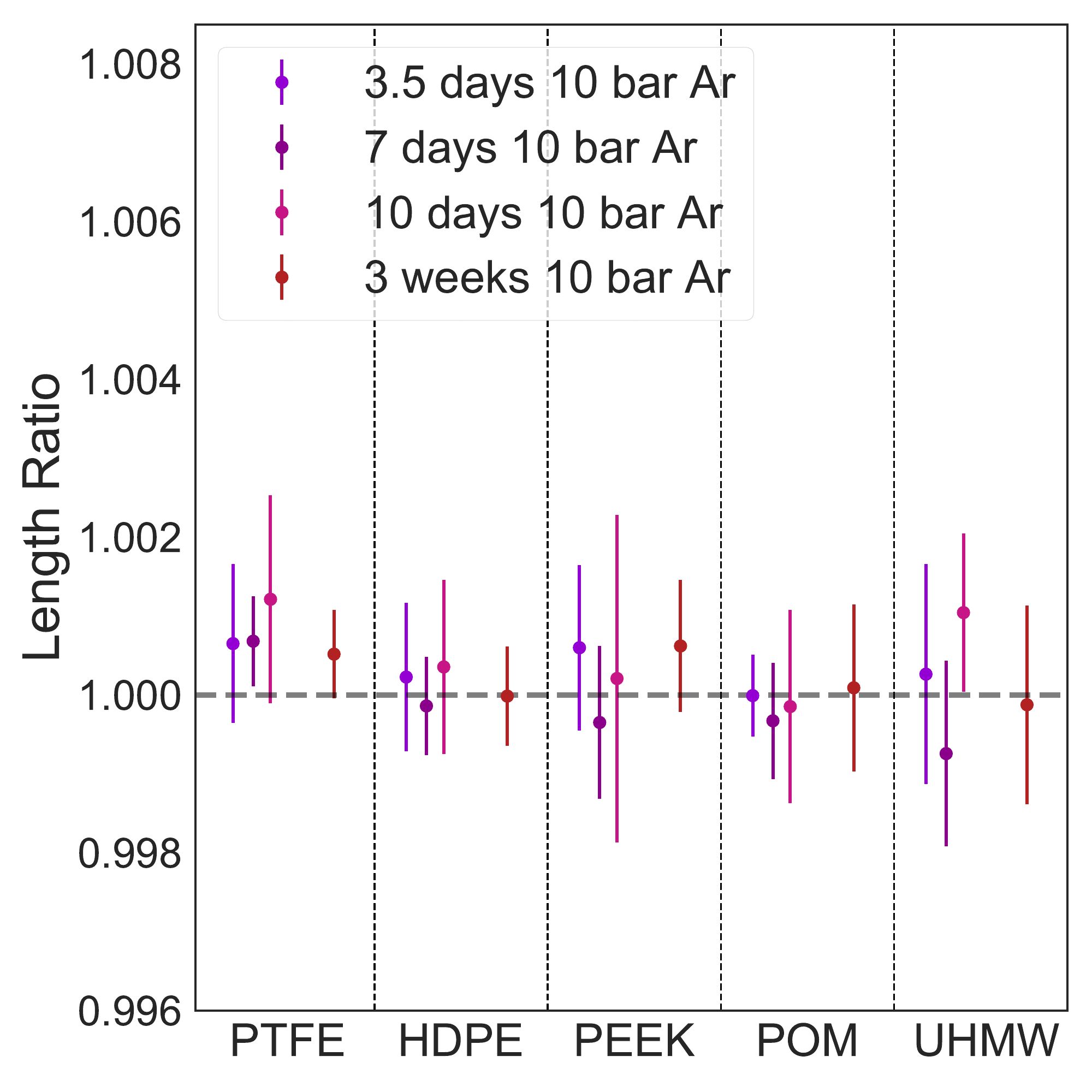}
\includegraphics[width=0.49\linewidth]{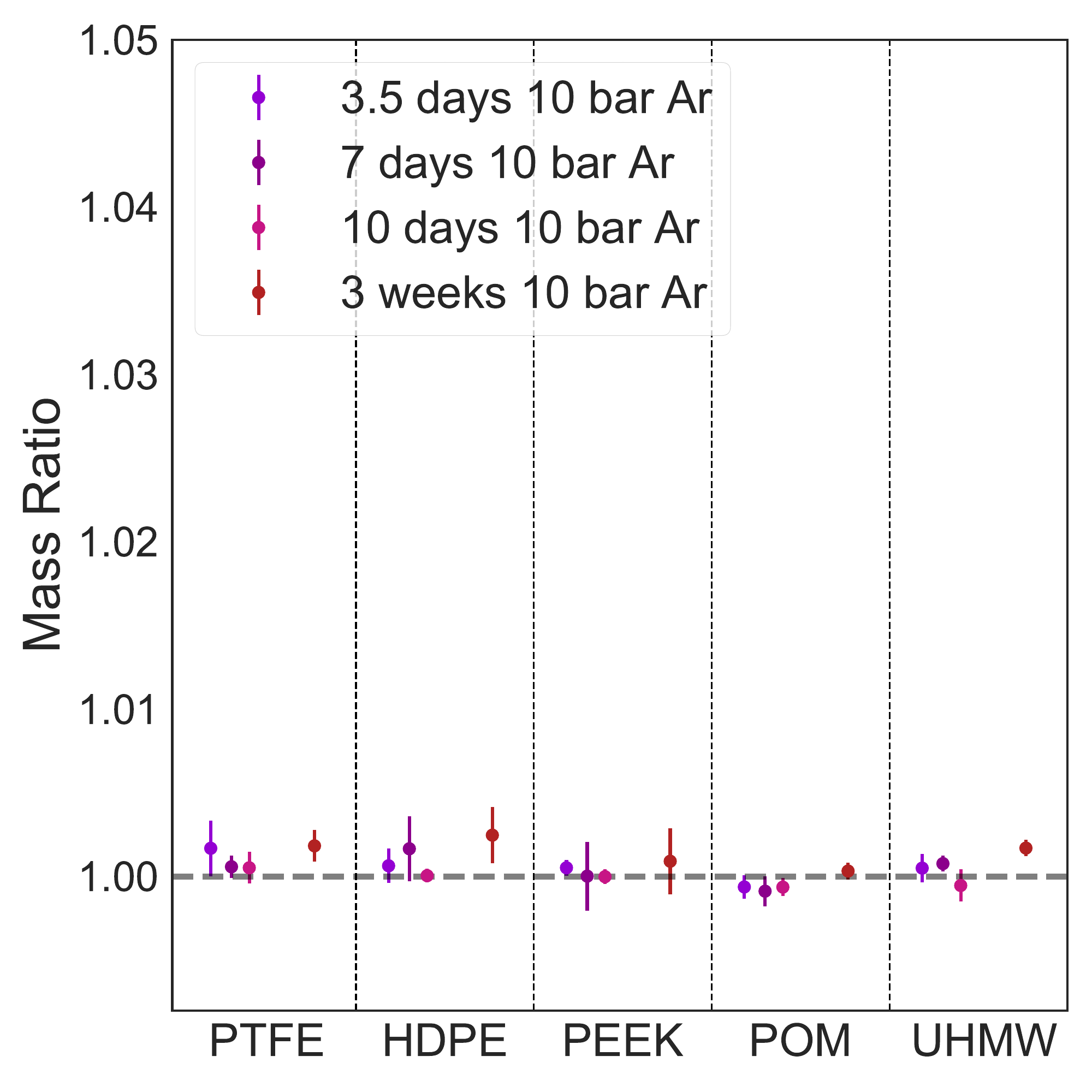}
\caption{Left: Ratios of lengths of posts after soaking for various lengths of time in argon at 10 bar divided by the original lengths of posts. Right: Ratios of masses of posts after soaking for various lengths of time in argon at 10 bar divided by the original masses of posts.  }
\label{fig:ArSwelling}
\end{figure}

\begin{figure}[t]
\centering
\includegraphics[width=0.49\linewidth]{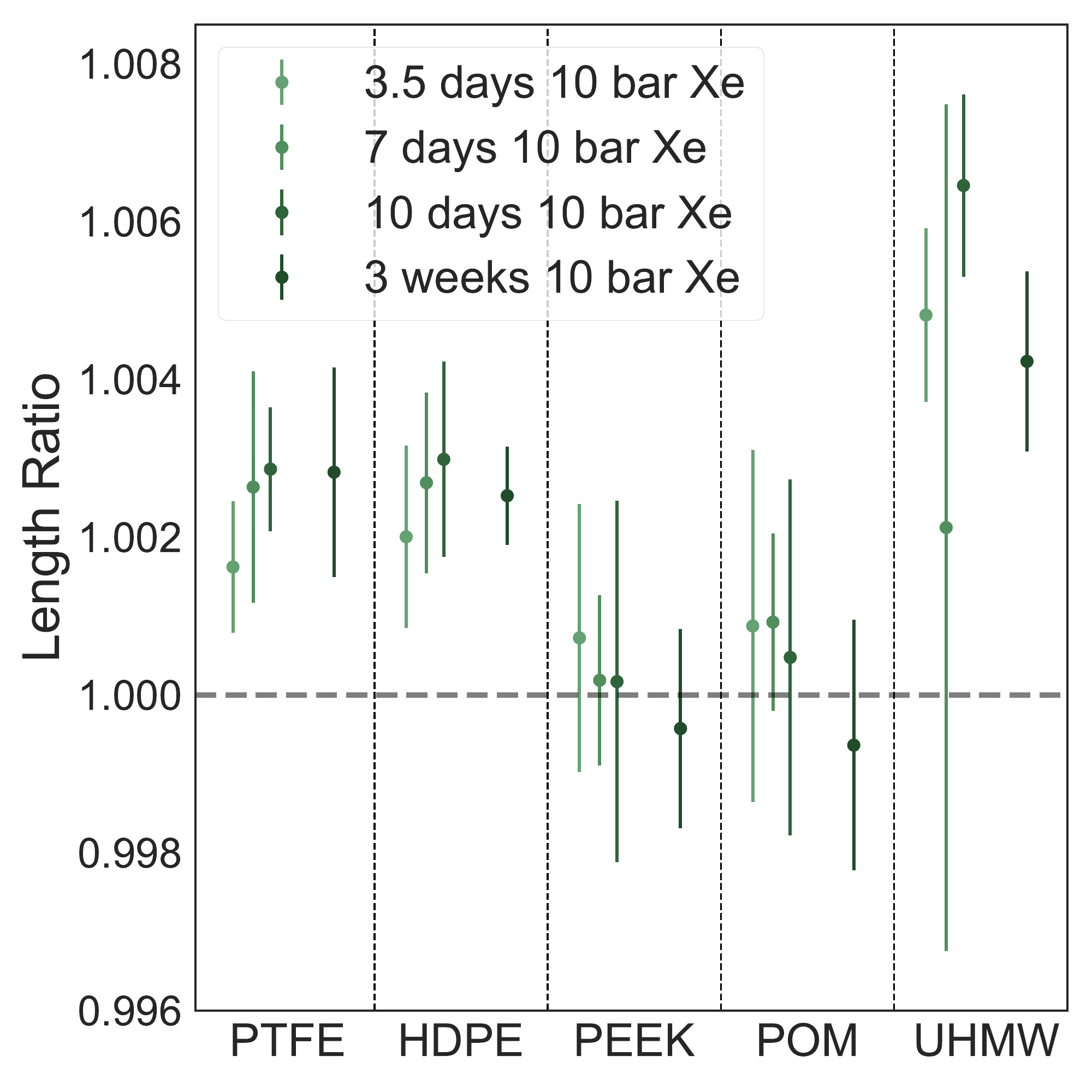}
\includegraphics[width=0.49\linewidth]{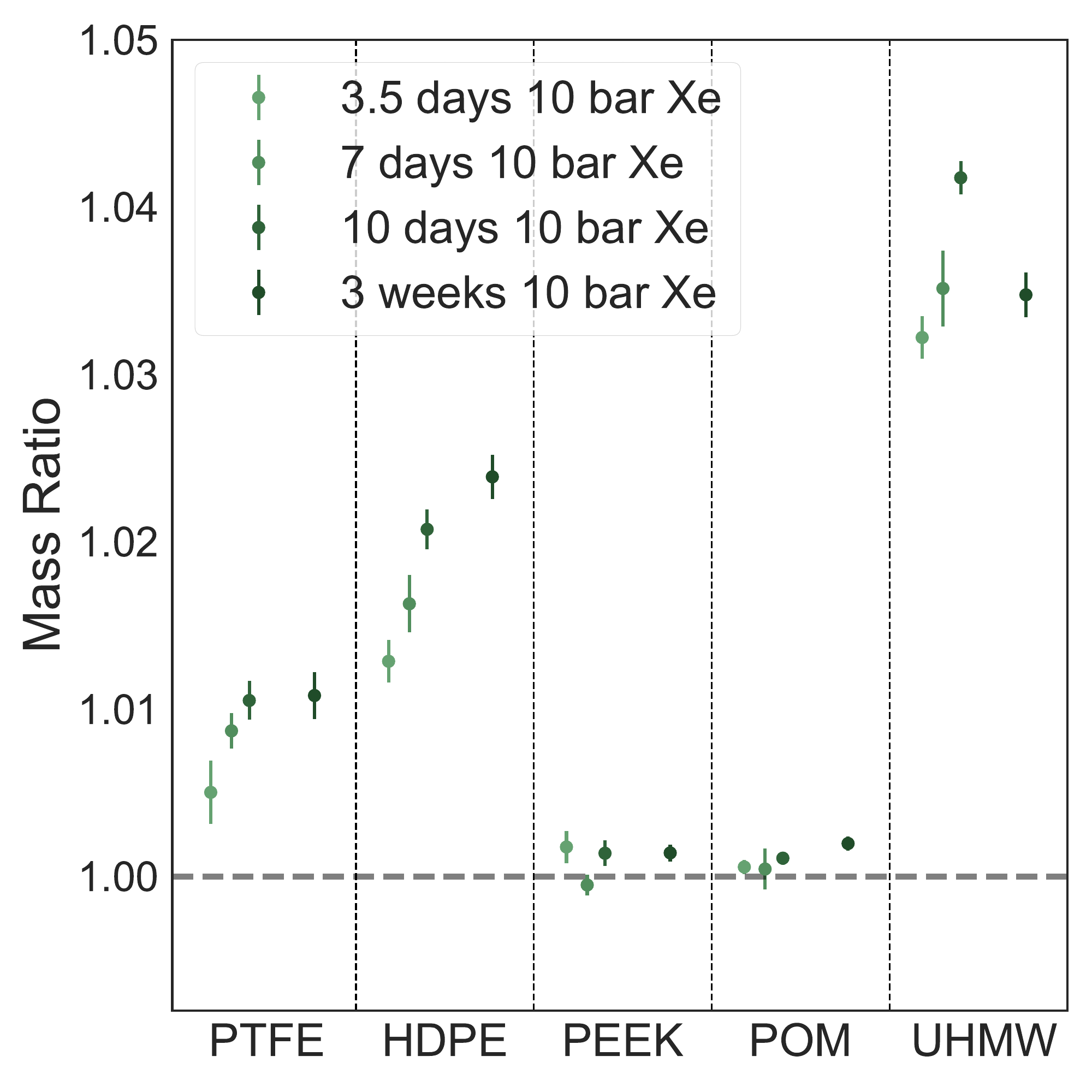}
\caption{Left: Ratios of lengths of posts after soaking for various lengths of time in xenon at 10 bar divided by the original lengths of posts. Right: Ratios of masses of posts after soaking for various lengths of time in xenon at 10 bar divided by the original masses of posts.  }
\label{fig:XeSwelling}
\end{figure}

Fig.~\ref{fig:ArSwelling} shows that no swelling was observed in argon for any material at the level of precision of this study.

Length and mass measurements in xenon are shown in Fig.\ref{fig:XeSwelling}. PEEK and POM were not observed to swell within the precision of this experiment, whereas PTFE, HDPE, and UHMW showed an increase in both mass and length. UHMW had the largest mass and length increase with HDPE and PTFE less. All three materials showed a continuing upward trend in their mass increases up to ten days.

\begin{figure}[t]
\centering
\includegraphics[width=0.49\linewidth]{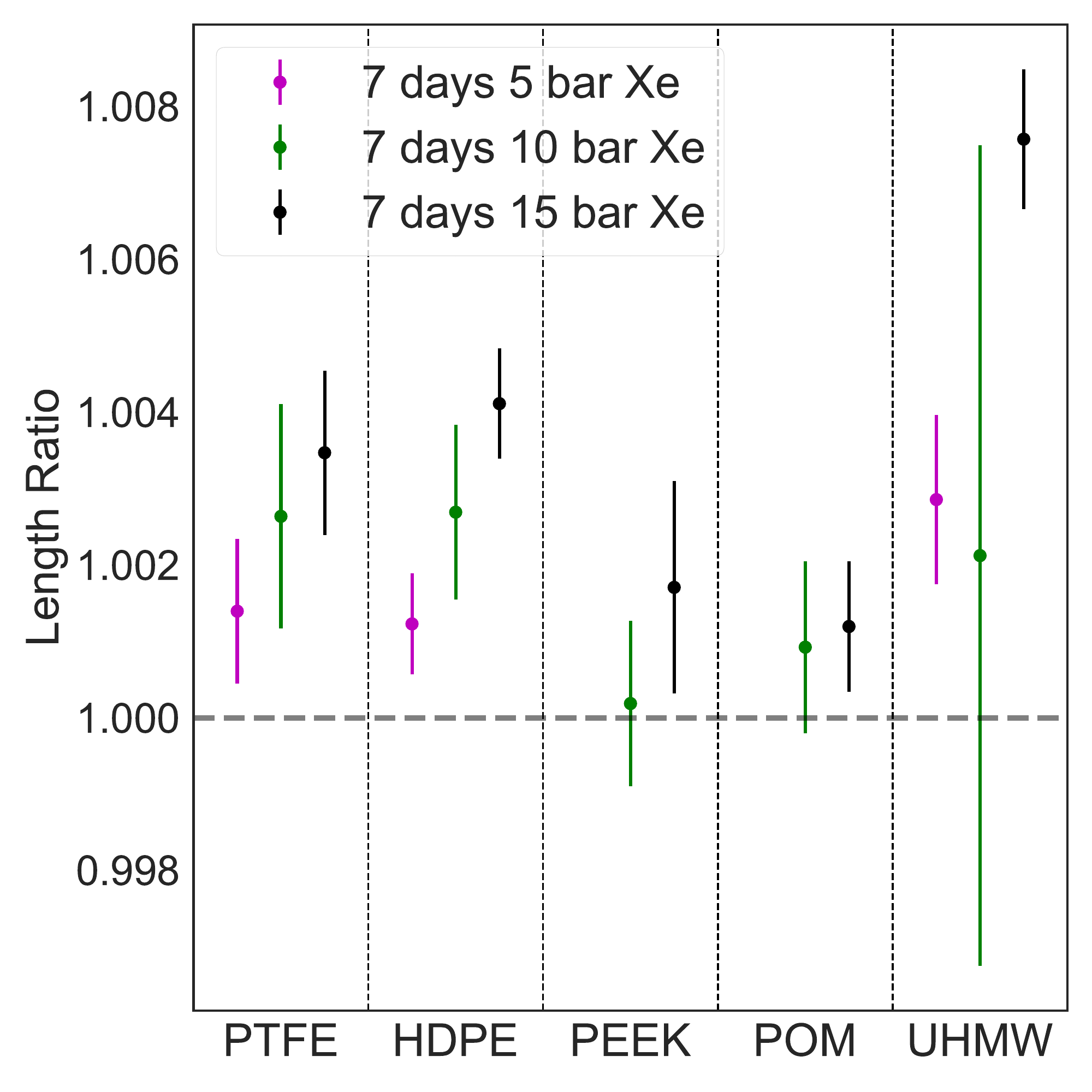}
\includegraphics[width=0.49\linewidth]{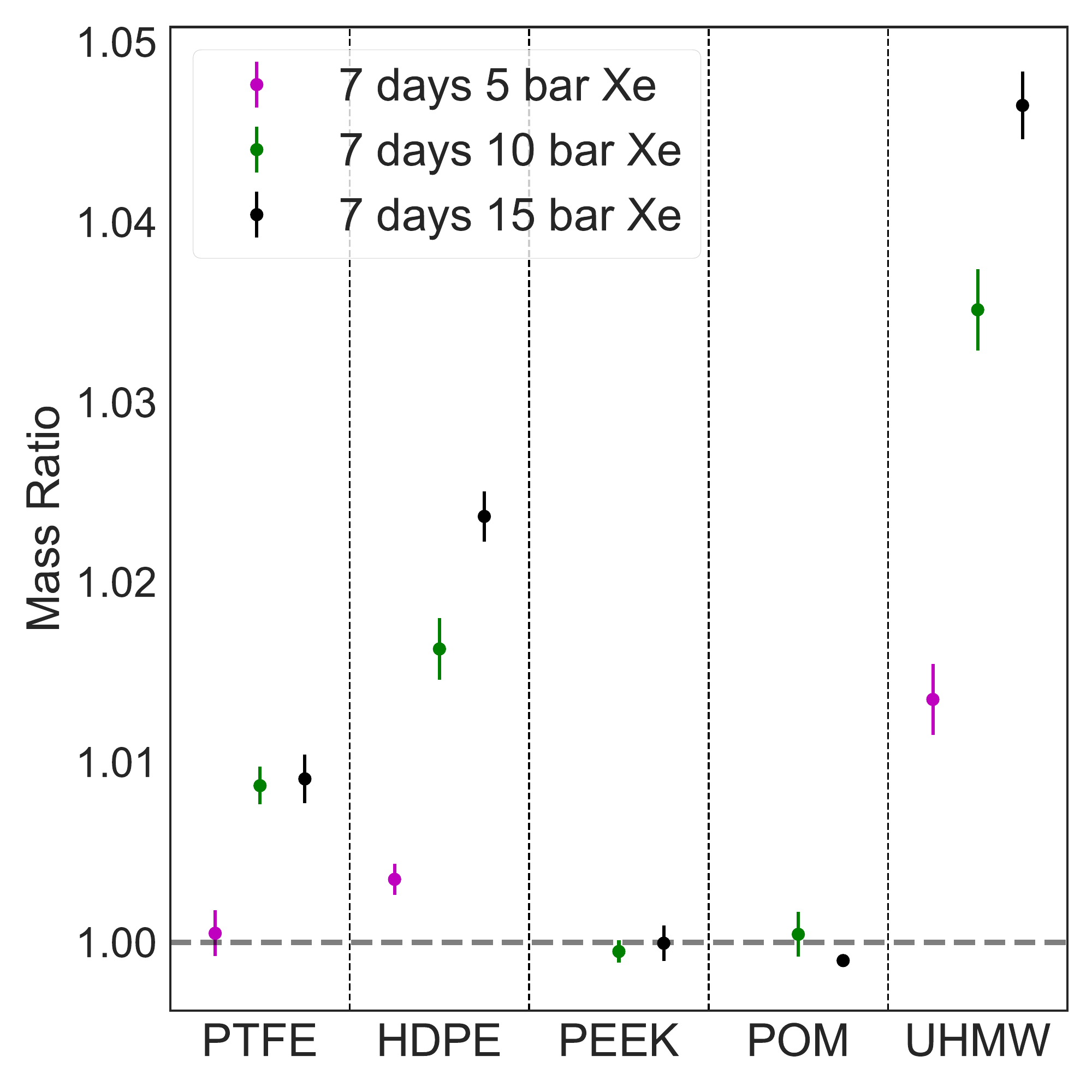}
\caption{ Left: Ratios of lengths of posts after soaking for a week in xenon at various pressures divided by the original lengths of posts. Right: Ratios of masses of posts after soaking for a week in various pressures of xenon divided by the original masses of posts.}
\label{fig:PressureSwelling}
\end{figure}

Higher pressure environments were observed to produce more absorption and swelling than the lower pressures within the same time period shown in Fig.~\ref{fig:PressureSwelling}. PEEK and POM did not have a perceptible change at any pressure.  The large error on the length of UHMW reflects the difficulty of measuring the length accurately, given that the material is soft and flexible.   This material experienced the largest gas absorption and length increase, absorbing almost 5\% of its mass in xenon in 10 days at 10 bar, and swelling in length by nearly 1\%.

\begin{figure}[t]
\centering
\includegraphics[width=0.49\linewidth]{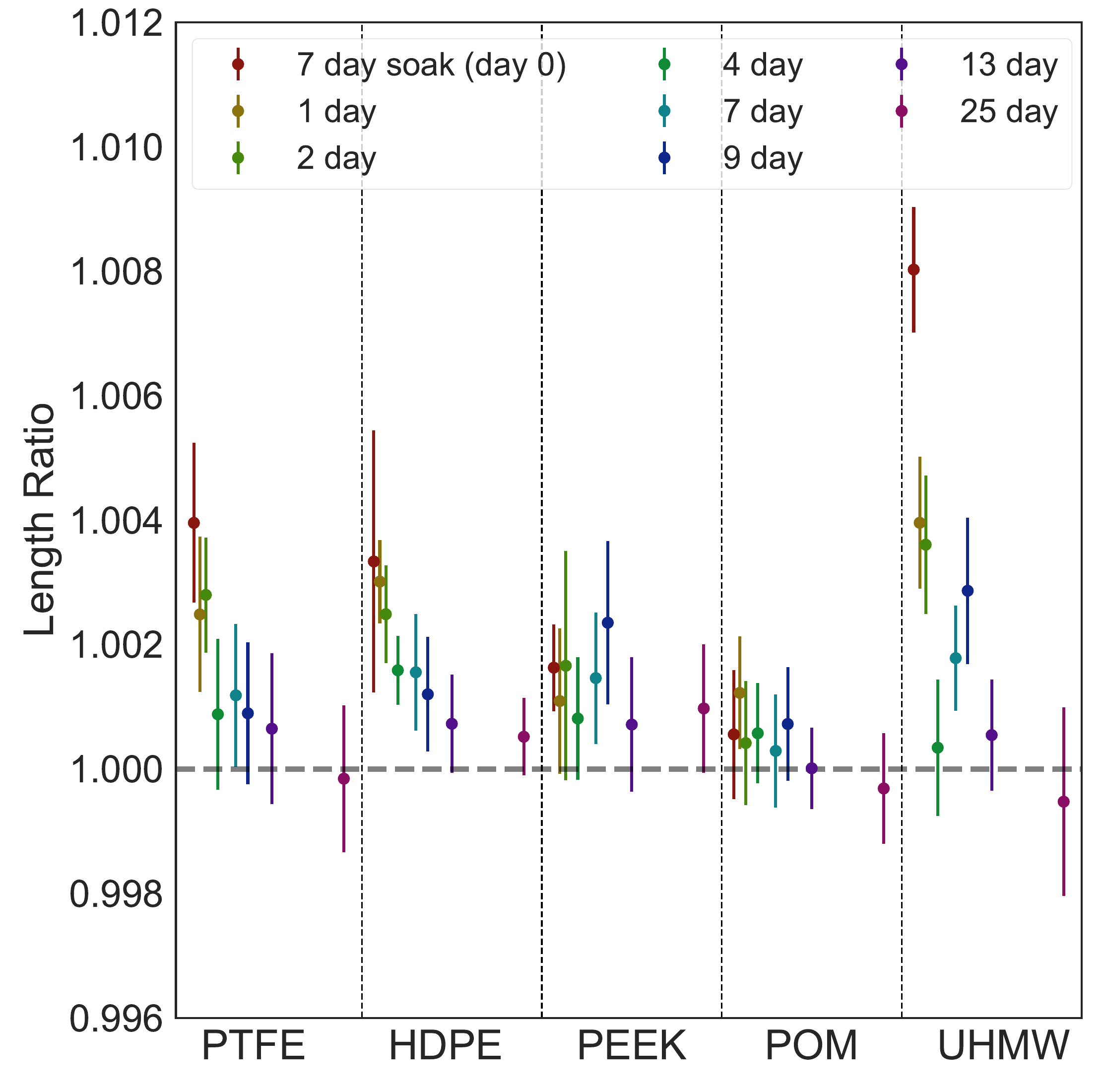}
\includegraphics[width=0.485\linewidth]{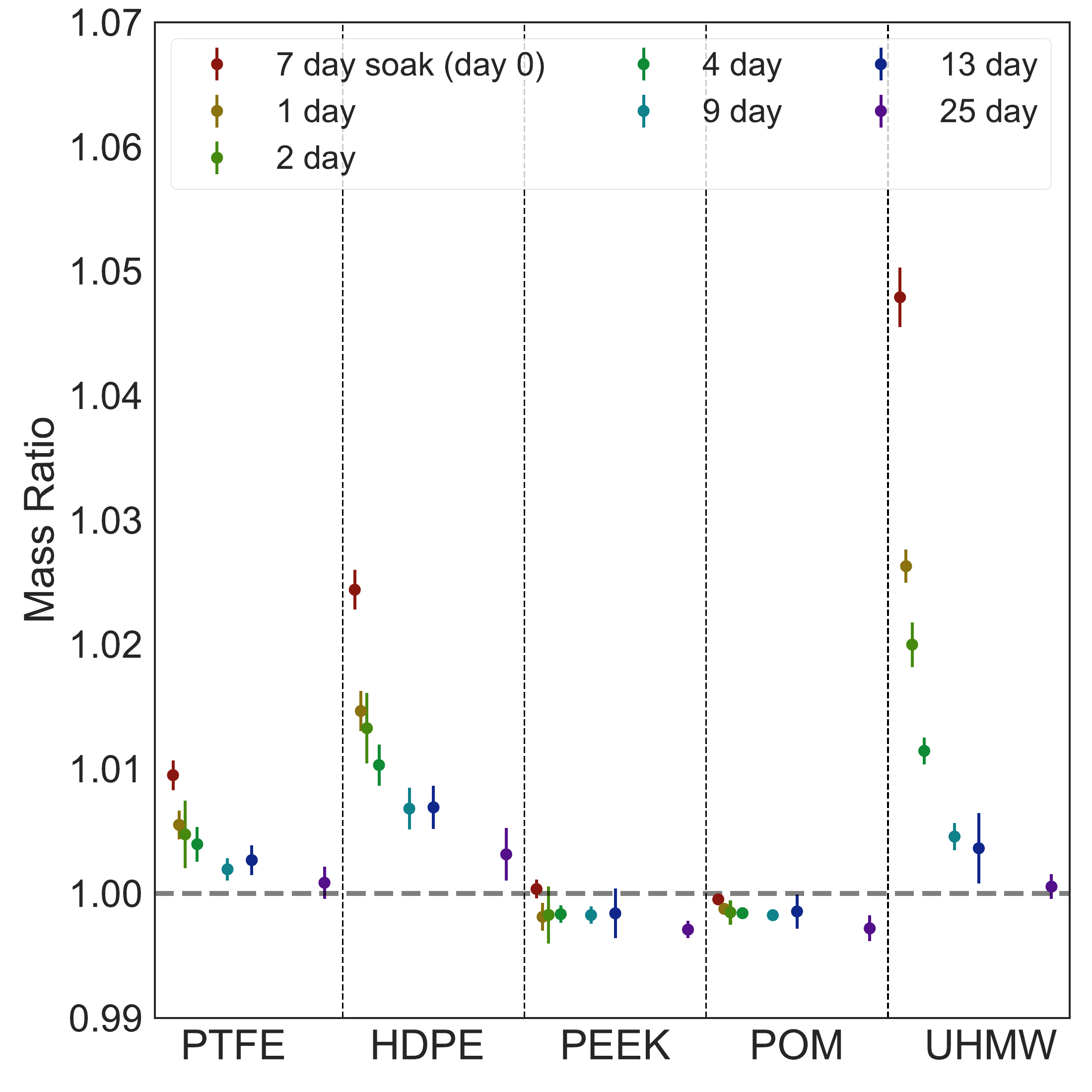}
\caption{Left: Ratios of lengths of posts after soaking for one week in xenon at 15 bar and then put under vacuum divided by the original length of posts. Right: Ratios of masses of posts after soaking for one week in xenon at 15 bar and then put under vacuum divided by the original, pre-vacuum masses of posts. }
\label{fig:Vacuum}
\end{figure}

Fig. \ref{fig:Vacuum} shows the effects of vacuuming materials after soaking in xenon at 15 bar for 7 days. As expected, the masses and lengths shrink as the xenon is drawn out of the material.  A ratio of 1.0 represents if the materials had no permanent swelling once they had absorbed and lost xenon - i.e. a return to the original pre-soak length. The lengths return to within 0.1\% of their original values, and masses to within .3\%, for all materials after two weeks of evacuation. After 25 days of evacuation all posts were consistent with their initial states.

\section{Tests of electrical strength of surfaces under high voltage \label{sec:Electrical}}
\subsection{Methodology}

\begin{figure}[b]
    \centering
    \includegraphics[height=0.4\linewidth]{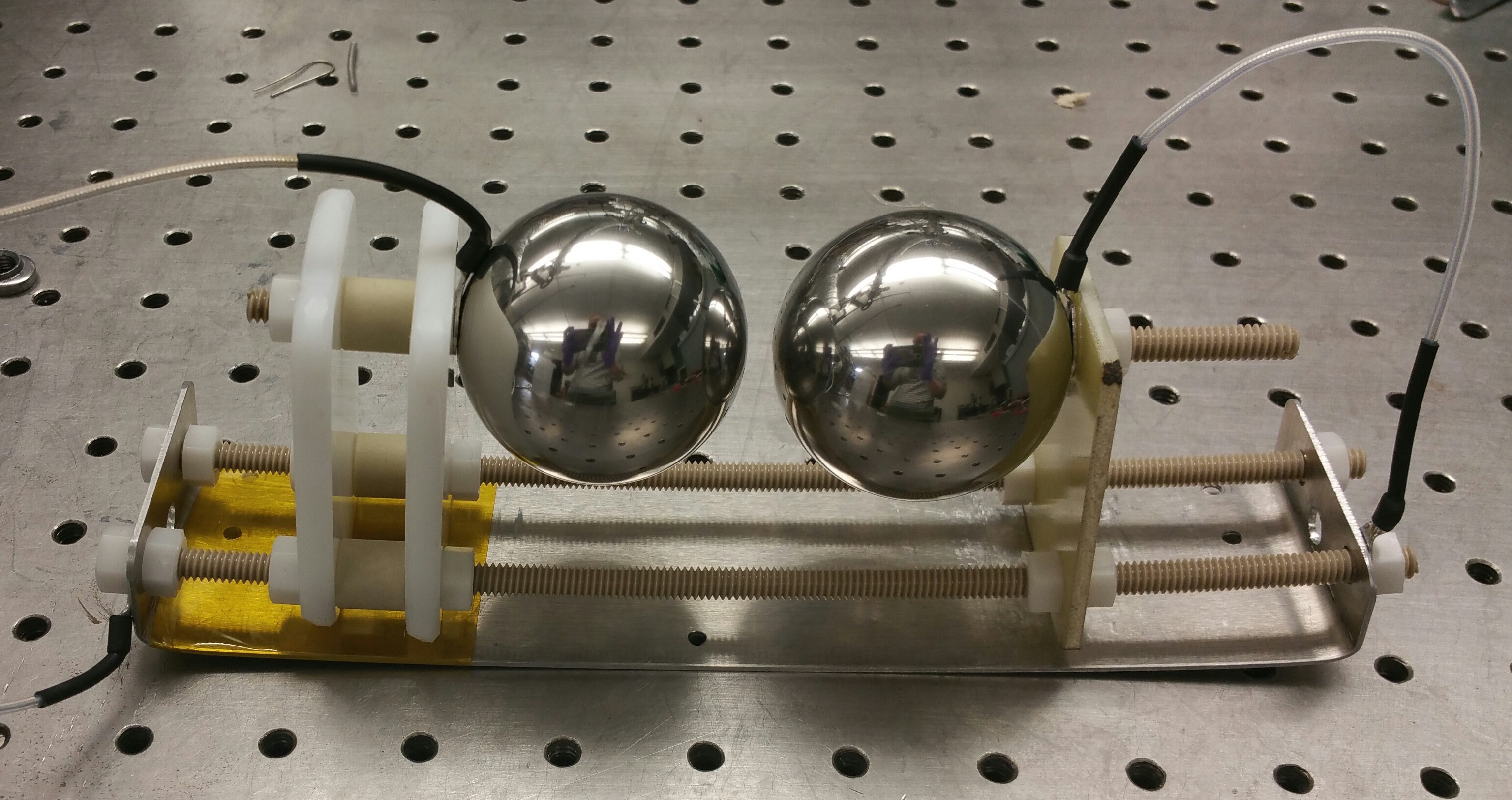}
    \includegraphics[angle=90, height=0.4\linewidth]{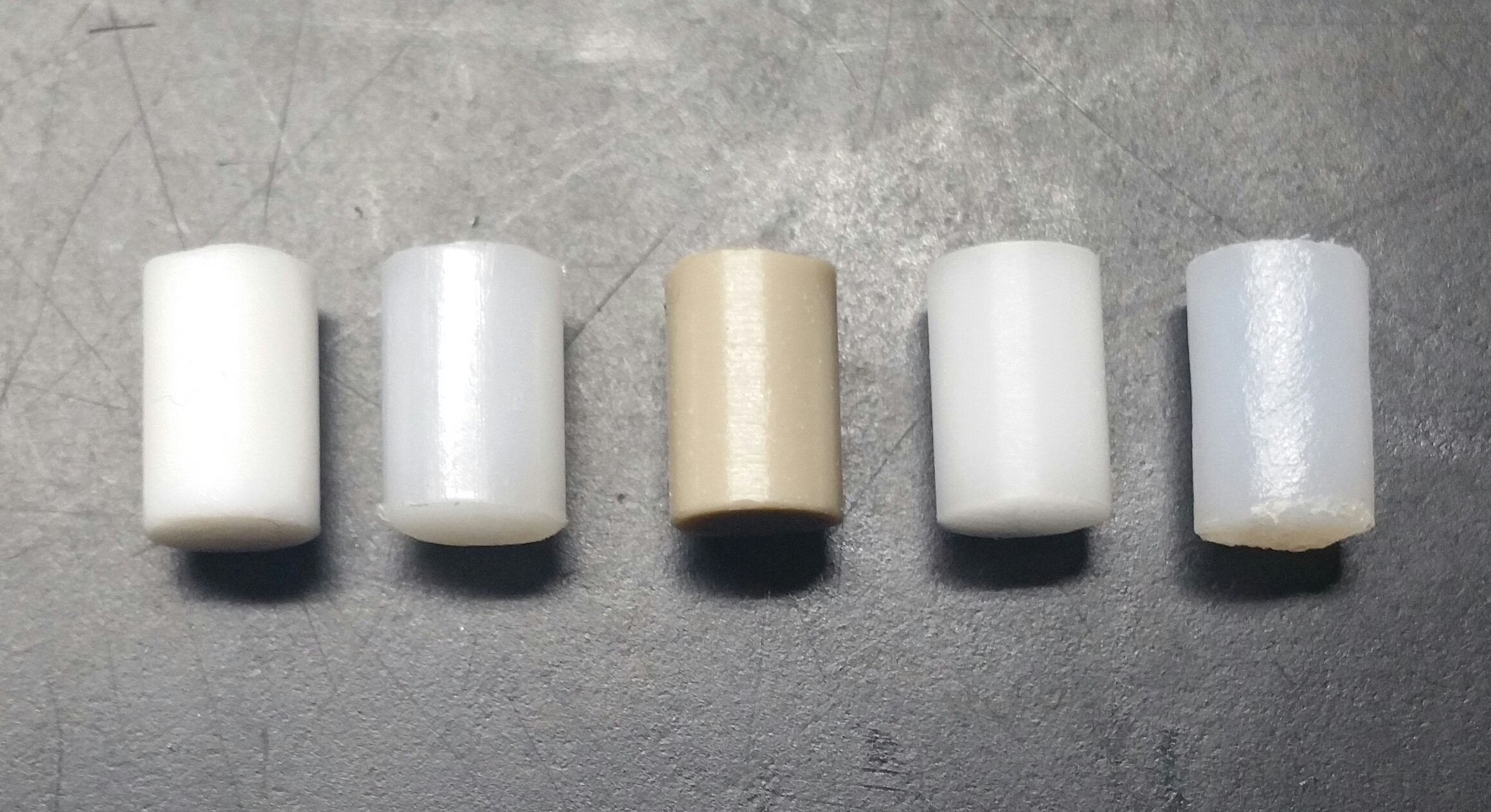}
    \caption{Left: Setup for holding the insulating posts in a smooth electric field where the left sphere is the high voltage, right is ground.  Right: Typical posts used for these tests. From top to bottom the materials are UHMW, POM, PEEK, HDPE, and PTFE.}    
    \label{fig:MagPostSetup}
\end{figure}

To test the breakdown voltage across the insulating materials, materials were cut on a band saw, sanded down to size and then deburred. The rods were cleaned for 15 minutes in ethyl-alcohol placed in a sonic bath before being exposed to high voltage. As a preliminary test to help rule out poor HV materials, 10 mm length posts of 6.4 mm diameter were placed between two spherical electrodes and the voltage raised until sparking occurred between the electrodes in air. All materials passed this preliminary scan except for antistatic UHMW, as described in Sec.~\ref{sec:Polymers}

The setup used to take these data can be seen in Fig.~\ref{fig:MagPostSetup}, where the right stainless steel ball bearing is connected to the chassis ground. The ground electrode is held with a threaded PEEK rod penetrating a G10 frame. The HV electrode ball is supported by a threaded PEEK rod penetrating two HDPE plates, separated with a ceramic spacer to lend extra structural strength. To secure an electrical connection, the ends of the electrical cables were fitted with panduit lugs, then attached to the PEEK rods between the electrodes and support plates. Nylon nuts were then threaded on and tightened to hold them in place. Both 51 mm ball bearings are tapped and drilled 12 mm deep for the PEEK rod to securely thread into and the opposite side ground down a slight amount to provide a flat spot for the insulating material posts to rest. %The supporting gas system is shown in \ref{fig:gaspanel}.

Several preliminary frames were tested to hold the ball bearings before settling on this design, which we denote A-C. Frame A used drilled acrylic blocks to hold the electrodes with PEEK rods, but the weight of the steel balls caused the acrylic to warp when heating the vessel during the vacuum stage, leading to misalignment of the spheres. The acrylic also outgassed profusely, leading to a high reading of water contaminants on the residual gas analyzer (RGA). Frame B was a high temperature 3D printable plastic, so that the spheres could be held without drilling into them, and Frame C was thin sheets of G10 with holes drilled through for the PEEK rods. Both Frame B and C gave sporadic values of breakdown voltage, oftentimes corresponding to milliamps of current being drawn and sometimes shorting of the system. The G10 showed visible carbon tracks along it, illustrating that the plates holding the high voltage sphere led to breakdowns when too insulative, likely due to charge pile-up effects causing a nonuniform electric field, leading to electrical breakdowns in places other than between the electrodes.  These preliminary studies informed final design with HDPE on the high voltage side and G10 by the ground as shown in Fig. ~\ref{fig:MagPostSetup}.

%\begin{figure}[t]
%    \centering
%    \includegraphics[ width=0.45\linewidth]{img/gaspanel.jpg}
%    \caption{Gas panel used to fill the vessel with gas with the bottom left two canisters being where gas is recaptured. }    
%    \label{fig:gaspanel}
%\end{figure}

After preliminary tests in air we placed the setup inside a 6 liter, 14.6 cm ID pressure vessel large enough to avoid the electrodes sparking to the inside walls. HV was supplied via a Glassman KT100-R20 power supply via an epoxy-potted and pressure-rated feed-through.  The feed-through was rated for 20~kV DC voltage, but a prototype was tested to 40~kV before this study.  If breakdown had not occurred by 30~kV, the voltage was decreased again.  For 5~mm posts, this is around three times the voltage specification required for the NEXT-100 EL region to obtain 2 kV cm$^{-1}$bar$^{-1}$.

%CITE https://www.lesker.com/newweb/gauges/ionization_kjlc_354.cfm
Once each post was between the electrodes and placed in the pressure vessel, the system was evacuated with a PFEIFFER HiCube 80 vacuum pump and baked at 85 degrees Celsius overnight.  This temperature was high enough to bake most of the water out of the system as checked by an RGA, but not so high as to melt the HDPE support plates. After cooling to room temperature a digital ion gauge \cite{IonGauge} was used to read a resulting vacuum  between 4x\(10^{-6}\) and 3x\(10^{-5}\) Torr. The vacuum line was then closed and the vessel filled with xenon gas. The voltage was then slowly increased between the spheres until there was a spark which could be recognized by an audible sound from within the vessel, and sometimes a current draw. The breakdown voltage varies with pressure, so each material was measured at varying pressures in 0.5 bar increments.  

%statement of how the RGA values varied with time, not just picture showing water

\begin{figure}[t]
\centering
\includegraphics[width=0.49\linewidth]{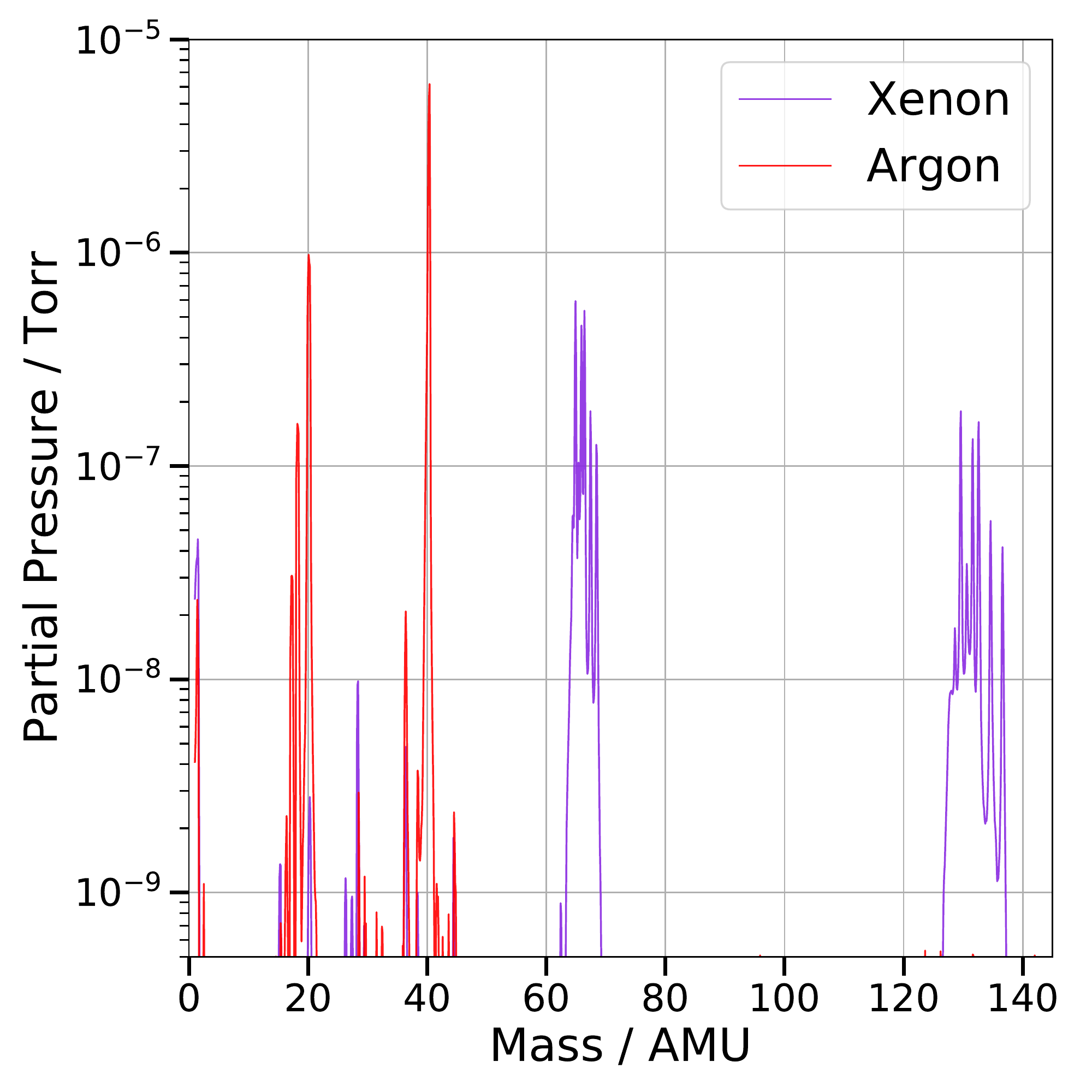}
\includegraphics[width=0.49\linewidth]{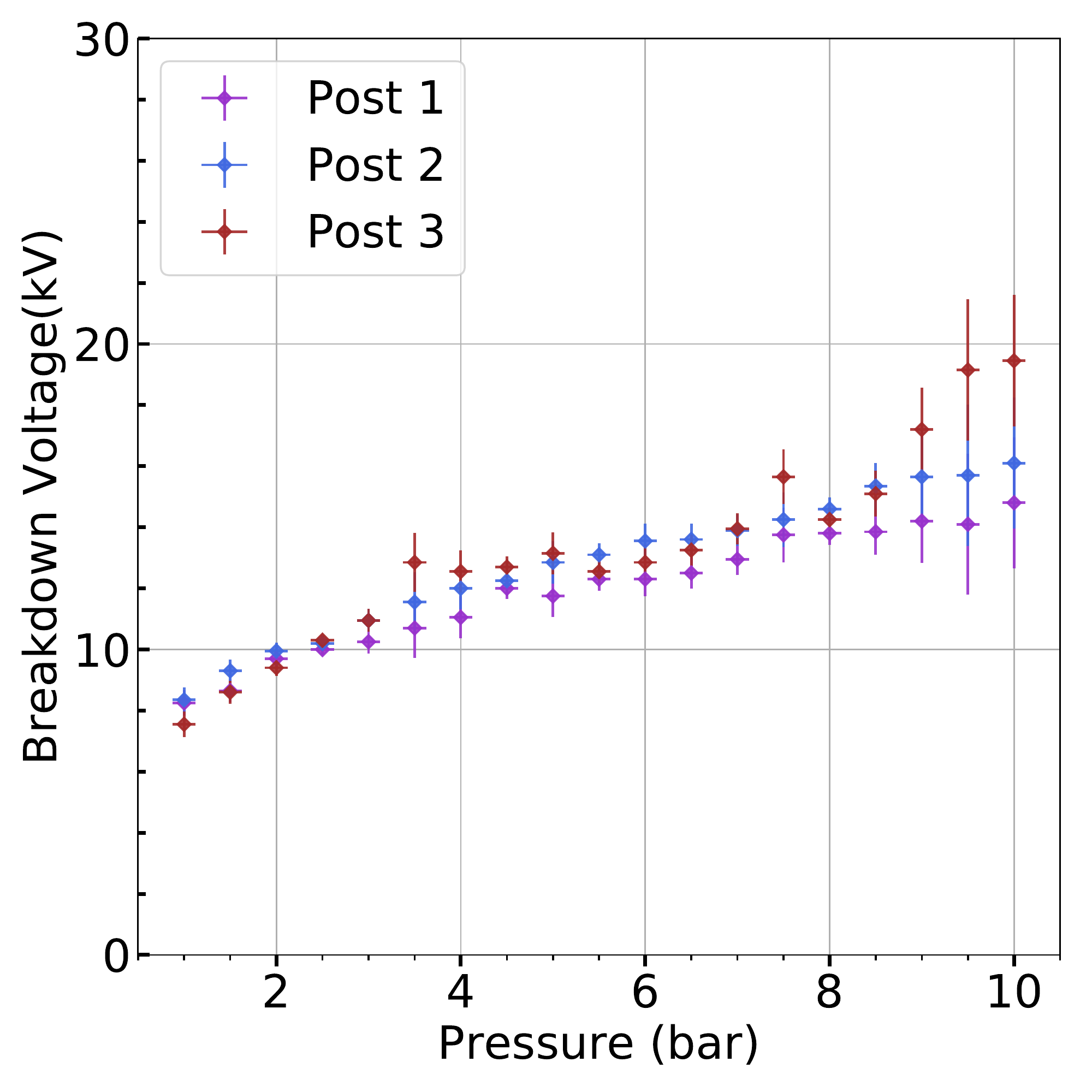}
\caption{Left: Reading from RGA at the end of a xenon and argon run with the vacuum subtracted resulting in peaks where we expect them for the two gases with very minor impurities. Right: Three runs of different HDPE posts under identical conditions showing that the data is repeatable. }
\label{fig:RGA}
\end{figure}

To ensure that the gas was not contaminated, an RGA scan was done at the beginning and end of every data run, obtaining results as shown in Fig.~\ref{fig:RGA}. The RGA plots show the atomic mass unit of the gases in the vessel. There were always peaks around 16 to 18, though these are consistent with the measured out-gassing from the vacuum lines alone. The left plot shows a reading from the end of an argon test, but with the vacuum scan seen as background and subtracted off. The double peaks for both gases is caused by the RGA double ionizing the atoms so that argon is shown at 40 and 20 AMU, and xenon is at 132 and 66 AMU. To check that this level of purity was adequate for reliable predictions of breakdown, the gas was circulated through a purifier (up to 10 bar), and tested again. The same breakdown voltage was achieved with the purified gas as with the gas directly from the bottle.

To test the repeatability of the system, a breakdown test was performed on three different 8.4 mm pieces of HDPE  in argon. Each post was placed inside the setup within the pressure vessel and evacuated overnight to 3x\(10^{-5}\) Torr, before closing off the vacuum and adding argon in half bar increments and checking for the break down voltage. The results of this test can be seen in Fig.\ref{fig:RGA}.

After taking the RGA measurements, the xenon was recaptured by cooling the recapture bottles in liquid Nitrogen to lower the temperature and therefore creating a pressure gradient to freeze the xenon into them. After this procedure, only a residual quantity of 300 millibar of xenon remained in the system, which was evacuated. The system was then filled with argon gas and the test repeated. After the data for argon were collected, the gas was vented out of the system down to 2 bar before closing off the gas panel and opening the vessel to switch out the post material. By running a test with a pre-mixed 3\% xenon-argon mixture and comparing with pure argon, it was shown that a small amount of xenon does not affect the measurements taken in argon.  This alleviates the possibility that residual xenon remaining after re-capture and evacuation influenced the argon measurements, which always followed with the same post shortly afterward. 
%percent the 3% xenon affected the measurements

All the materials that passed the preliminary test in air were further tested using a 5~mm length post in both xenon and argon. To compare to the breakdown of pure gas, breakdowns between the spheres at the same lengths as the posts were also investigated.
 
%Then the most promising material for an EL mesh support was run at 7.5 mm in both gases as well. 

\subsection{Results}

\begin{figure}[t]
\centering
\includegraphics[width=0.49\linewidth]{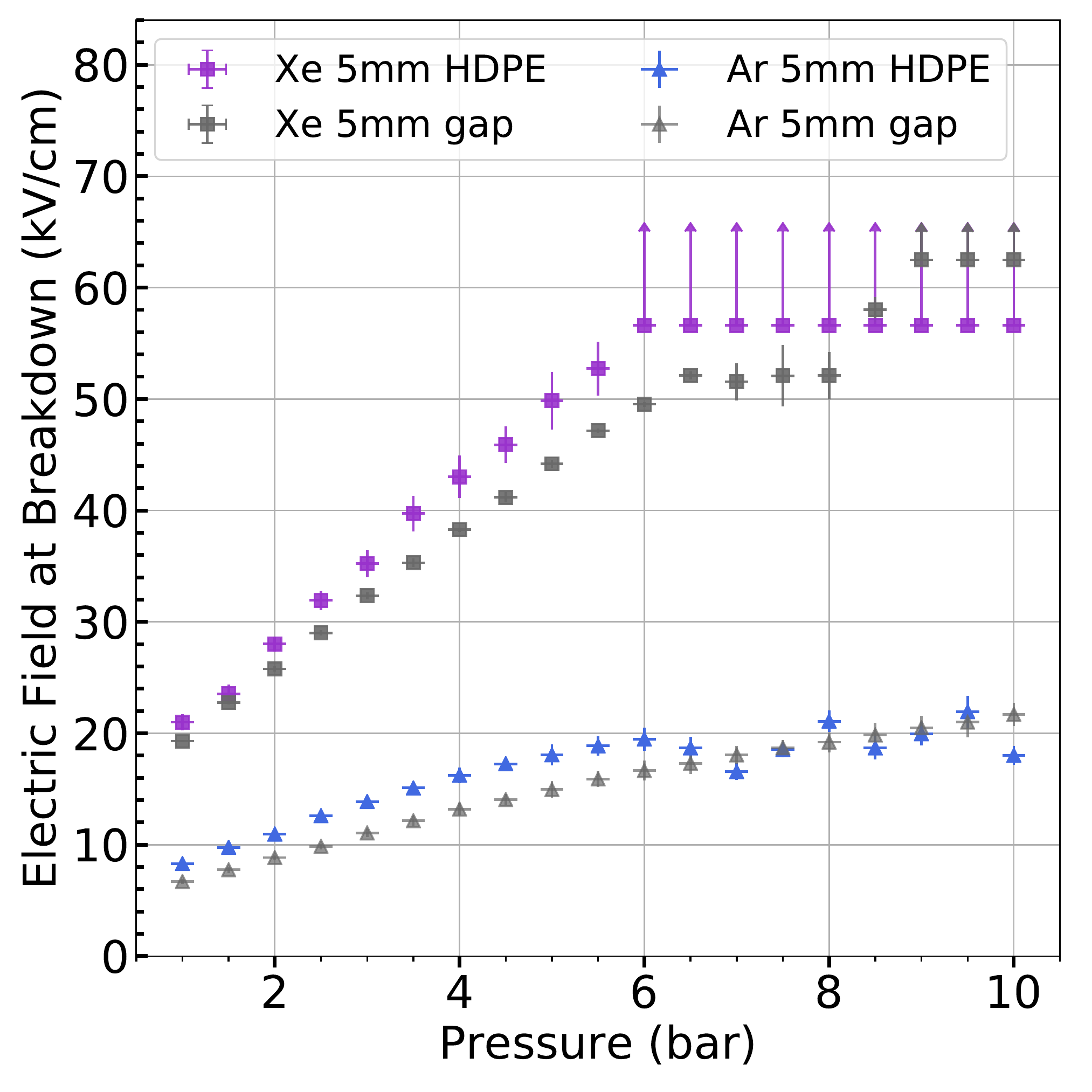}
\includegraphics[width=0.49\linewidth]{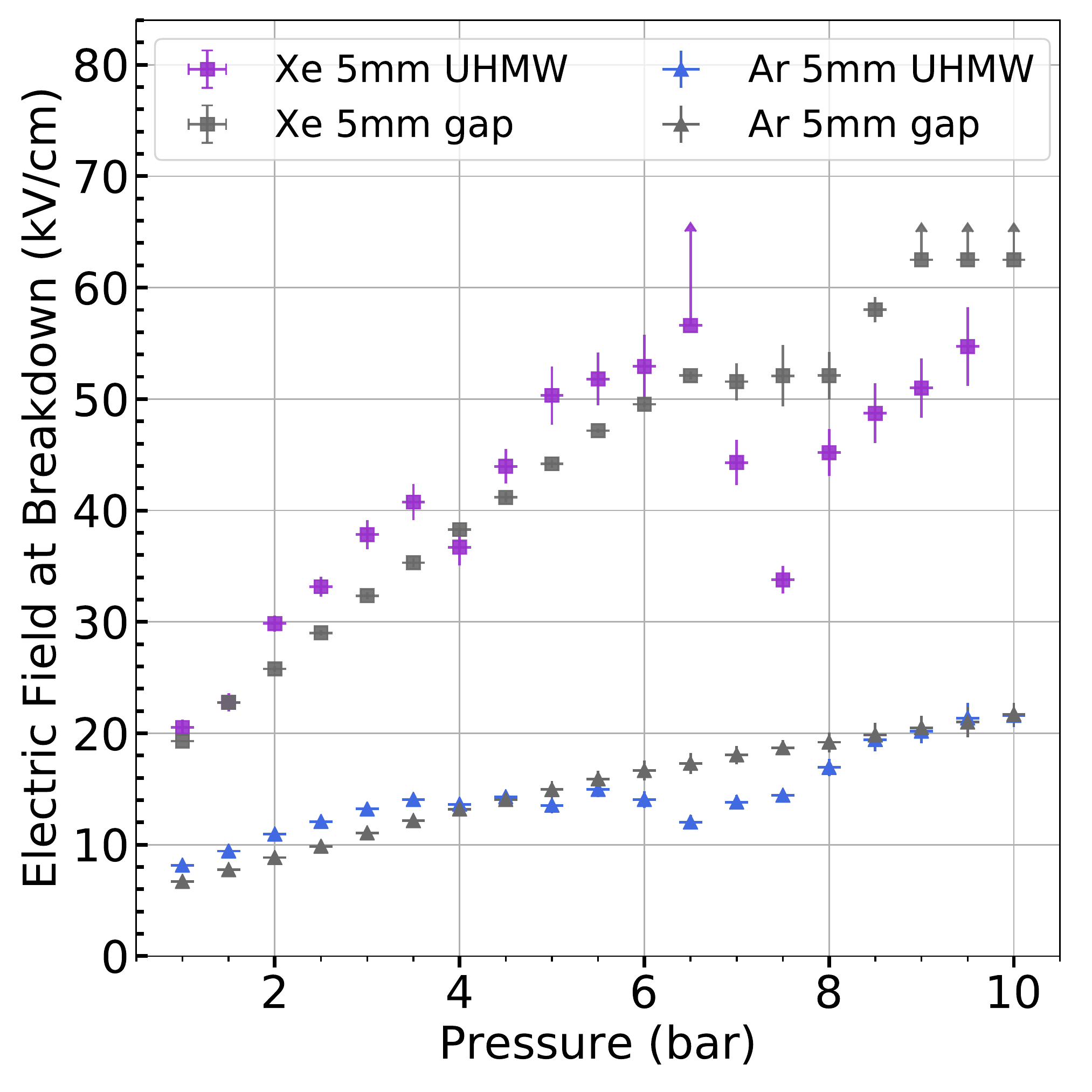}

\includegraphics[width=0.49\linewidth]{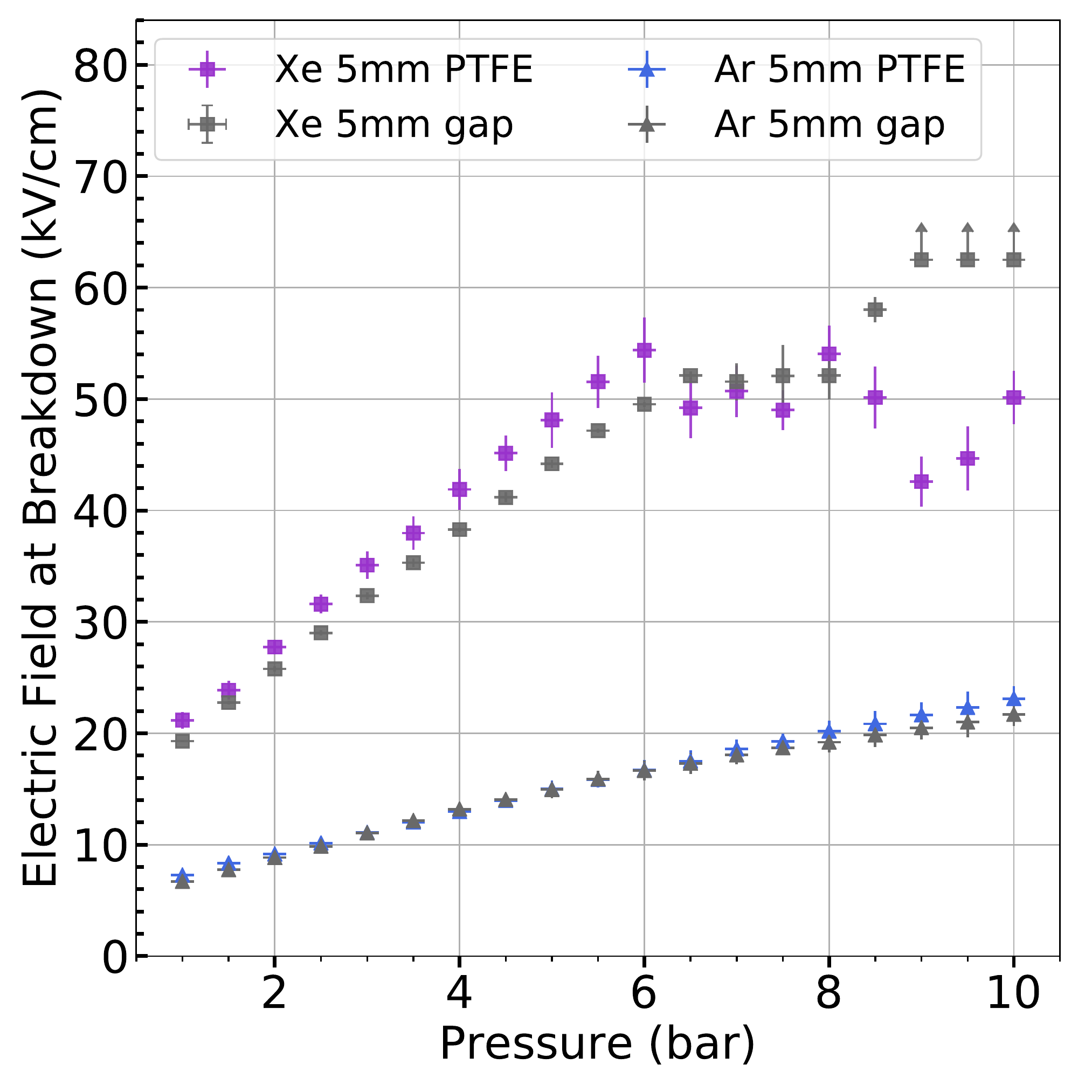}
\includegraphics[width=0.49\linewidth]{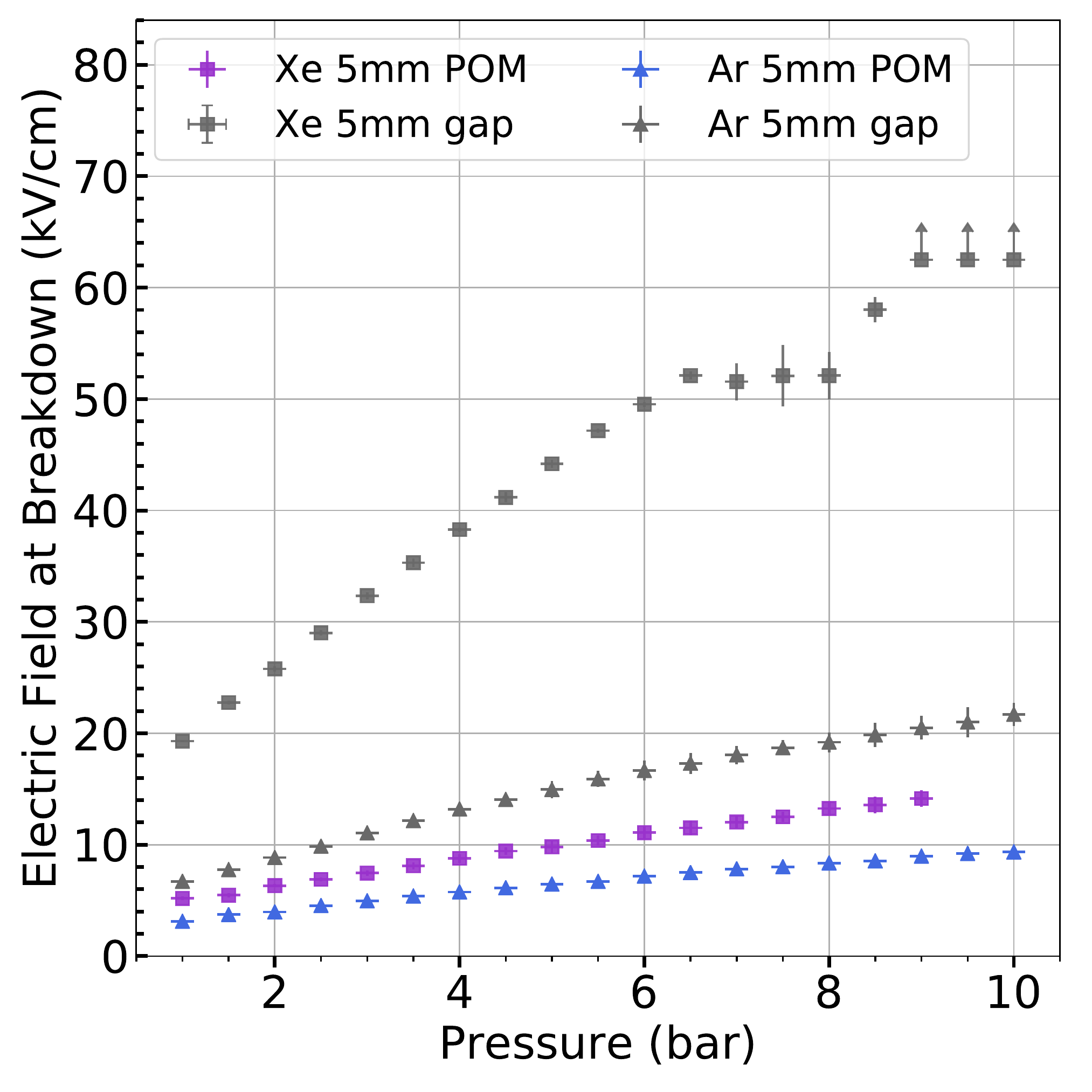}
\caption{Breakdown point of various materials at different pressures in xenon and argon gases, with the grey points being the breakdown through the respective noble gas with no material between the electrodes. The high voltage feedthrough was only rated for 20kV so we stopped taking data at 30kV to reduce the risk of breakage.  Upward arrows report that 30~kV was held without breakdown at this pressure.}
\label{fig:HVplots}
\end{figure}

Fig.~\ref{fig:HVplots} shows the breakdown voltages of 5~mm posts of each material as a function of pressure in pure xenon and argon gases.  PEEK is not shown, due to its destructive failure at the relatively low voltage of 12~kV at 4 bar in xenon gas. This burned a 4.1 mm track into the material and reduced its subsequent breakdown strength by almost 80 percent. This was the only catastrophic failure in any of the materials in these studies and is shown in Fig. \ref{fig:destruction}.

At 1 bar the xenon and argon curves match within 20 percent of Paschen predictions for pure gas with explanations for the discrepancy given in \cite{GONZALEZDIAZ2018200}. All of HDPE, UHMW and PTFE appeared to increase the maximum voltage held versus a pure xenon gap at pressures below 6 bar, likely due to a reduction in total stressed area between the electrodes. At higher pressures, both UHMW and PTFE did not continue to show this effect at higher pressures / voltages, instead becoming stochastic in their breakdown point, which may be interpreted as these materials enabling surface discharges.  HDPE maintained this strengthening effect even at the highest pressures.

In argon gas, all of HDPE, UHMW and PTFE demonstrated a similar breakdown strength to a pure argon gap. Thus in this case it appears the spark is propagating across the argon itself, rather than via a surface mediated effect. In both gases, POM had breakdown values less than half of pure gas alone and should be avoided inside detectors.

%\begin{figure}[t]
%\centering
%\includegraphics[width=0.49\linewidth]{img/LengDep.pdf}
%\caption{Electric field at the breakdown point for various lengths of HDPE and gaps in argon and xenon. The high voltage feedthrough was only rated for 20kV so we stopped taking data at 30kV to reduce the risk of breakage.}
%\label{fig:Efield}
%\end{figure}

%HDPE showed the highest and most consistent break down in xenon, so we tested how it behaved with different lengths. 

%Fig. \ref{fig:Efield} shows that breakdown has a direct electric field dependence in argon for HDPE and gaps, independent of gap length. Xenon shows the gap stays at relatively the same breakdown field at different lengths  The higher voltages with larger spacing in xenon appear less predictable than the breakdowns in argon and are discussed in the next section.

%and the HDPE provides equal electric fields with lengths.

\section{Discussion \label{sec:Discussion}}

We have characterized the absorption of noble gases into polymers at high pressure, and the electrical strength of various materials in high pressure argon and xenon gases.  While no swelling or mass increases are observed with argon gas, HDPE, PTFE and UHMW all exhibited substantial length and mass changes under exposure to xenon gas, with larger increases over longer times and at higher pressures.  This suggests a qualitative difference in the absorption behaviour in argon and xenon environments.

The diffusion of noble and other gases through polymers has been studied in various contexts, because its of relevance for industries including the oil and gas industry and food packaging.  References such as \cite{measurement}, \cite{pressuredep} and \cite{transport} report on the seepage of gases through very thin polymer membranes.  The effect on mechanical properties of the membrane, such as changes in shape or mass, are not reported.

Because they are heavier and thus move more slowly in thermal equilibrium, a naive expectation might be that more massive noble gases would diffuse through materials more slowly than less massive ones.  This is observed in only some conditions. For example, in \cite{Schowalter2010}, xenon permeability through Kapton films is shown to be a factor $\sim$10 times lower than krypton and $\sim$100 times lower than argon at the same temperature.  However, careful analysis of the time profiles and pressure dependencies of noble gas permeation as reported in Ref~\cite{pressuredep} illuminates more nuanced behaviour.  There, two distinct absorption mechanisms are inferred in oriented polypropylene (OPP) from the two time constants in absorption and desorption curves. These are named  the ``normal pathway'', or standard diffusion through the solid polymer matrix, and ``cavity condensation'', which occurs only for the heaviest gases and only at the highest pressures.  Under this mechanism, small numbers of atoms may enter material voids, where interactions with the walls create potential wells that increase the local effective pressure.  This causes condensation of the gas in these voids, enhancing the absorption rate. Related phenomena have been observed in porous carbon media, where enhanced production of nitrogen dimers is observed and effective pore pressures in excess of 1000 bar can be generated \cite{KANEKO1996319}. This process is postulated to also be active in the amorphous regions of polymer matrices, with measurements from \cite{pressuredep} suggesting that it accounts for 8\% of diffusion of xenon at 1 bar in OPP.  The characteristic two-exponential behaviour is reported for Xe and Kr but not Ar, He or Ne, which have much lower boiling points.

At the pressures considered in this work, the cavity condensation mechanism is expected to be further enhanced relative to that reported at 1 bar in  \cite{pressuredep}.  This may provide an explanation for the striking qualitative difference between xenon and argon absorption in bulk HDPE, UHMW and PTFE.  The difference between these materials and POM and PEEK would then be understood to lie in the different density of microscopic voids in these materials that may enable the condensation mechanism.

\begin{figure}[t]
\centering
\includegraphics[width=0.49\linewidth]{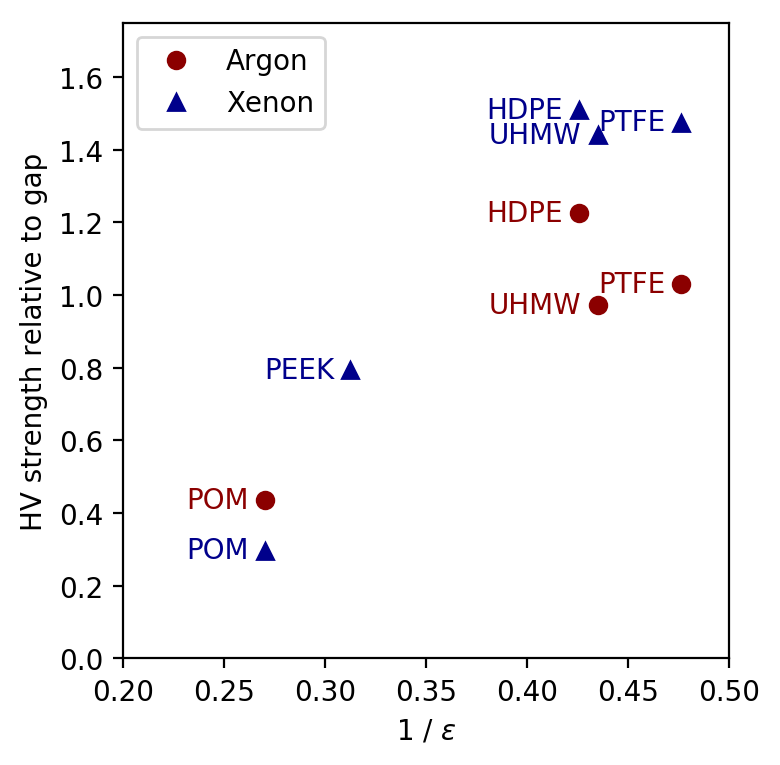}
\includegraphics[width=0.49\linewidth]{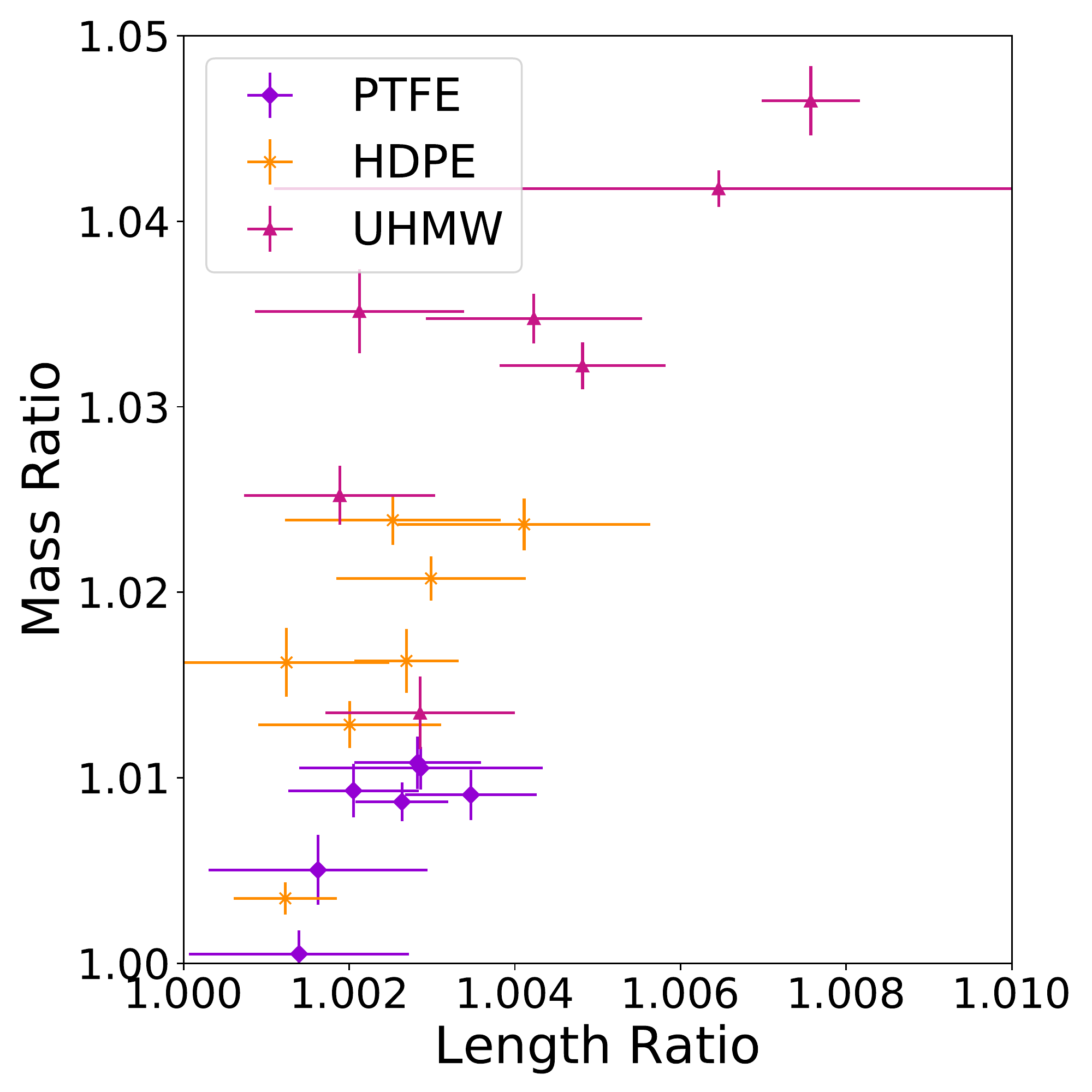}
\caption{Left: Comparison of breakdown strength at 5.5 bar with inverse dielectric constant (except PEEK, evaluated at 4 bar, before destructive failure). A correlation with dielectric constant is observed.
Right: Comparison of mass and length ratios of PTFE, HDPE, and UHMW soaked in xenon showing each material having a linear correlation between its mass and length. }
\label{fig:VsDielectric}
\end{figure}

Fig.\ref{fig:VsDielectric} (right) displays mass versus length for HDPE, UHMW, and PTFE soaked in xenon both with different pressures and different soaking times. There is a clear correlation between mass and length change ratio for each of the materials, with PTFE appearing to display a distinct slope compared to the other two. In the case of uniform absorption it would be expected that mass increases as the length increase cubed. Since this isn't the case it would appear the walls of the posts are expanding more than the length.

In Fig.\ref{fig:VsDielectric} (left) the dielectric strength of these materials show similar behaviours between high pressure argon and xenon gases. More scatter was observed between breakdown voltage points at different pressures in xenon.  This may be because dielectric strength of xenon is simply higher, and thus the relevant surfaces carry more charge which may fluctuate to initiate a breakdown.  This scatter is not observed for HDPE, which maintains a reliable and predictable breakdown strength until the pressure becomes sufficiently high that breakdown cannot be achieved in our system.  In both cases, POM is the weakest material, apparently carrying surface breakdowns at low voltages in both cases.

We noted that under higher breakdown fields, some pitting occurred in the electrodes, and a dark residue was observed to accumulate on the HV cable insulation and posts, which we suspect may be the ejected material from the spark.  This may potentially explain the more scattered behaviour in xenon gas, as this ejected material could introduce new but sporadic conductive paths.  To ameliorate this effect on our key conclusions, we always started with the highest pressures first and then worked toward lower pressures by re-condensing xenon gas into the bottle. Thus the highest voltage points use the most pristine setup and are most reliable.  A repeated run with HDPE, which did not show significant scatter or reduction in strength, suggests that there was no significant memory effect, producing an effectively identical set of data points.  A repeated run with PTFE also did not show evidence for hysteresis, but it is more difficult to be conclusive due to the wider spread of points at high pressures.

Prior work with HV insulators in liquid argon \cite{Lockwitz:2015qua} found that the threshold for surface breakdowns was inversely proportional to the material dielectric constant. This phenomenon has the intuitive explanation that materials with high permittivities induce larger electric fields near the interfaces which may initiate breakdowns. We observe approximately this trend in both high pressure argon and xenon gases (Fig.~\ref{fig:VsDielectric} left). In this plot, the strengths are shown relative to an empty gas gap at 5.5~bar, which is below the regime where large point scatter was observed.  For PEEK, which failed destructively at 4~bar in xenon, the ratio is shown at this pressure.

\section{Conclusions}
In this paper, we show that argon gas does not cause swelling in PTFE, HDPE, PEEK, POM, or UHMW. However, in those same materials PTFE, HDPE, and UHMW do absorb Xenon, with over 4\% mass increase in UHMW over 10 days and a more moderate 2.5\% mass increase in HDPE after 3 weeks. We have also shown that a variety of polymers can raise the voltage required for sparking over a distance in gas.  In both xenon and argon, HDPE appears to be the strongest and most stable option near high electric fields as it has a consistently higher breakdown voltage. 

For structural supports within an EL region, the critical property is the ability to withstand a high electric field without sparking. In this work we have demonstrated that the materials such as HDPE and UHMW appear well suited for this purpose. Without further information, HDPE seems the optimal material as it consistently strengthened the gap's ability to hold high voltage relative to pure xenon, and had a consistent and predictable breakdown behaviour even at the highest pressures. Based on these data, HDPE has been preliminarily selected for the EL region support material to be used in the NEXT-100 and NEXT-ton detectors.

As expected, we find generally higher dielectric strengths in xenon than in argon. We conclude that the practice of testing components for HV strength in argon gas before installation in xenon detectors appears valid.  Our measurements support the dependence of the insulator on inverse dielectric constant as has been shown in other works.

The length changes observed in PTFE, UHMW and HDPE under Xenon exposure have implications for the tolerances that can be achieved using these materials in Xenon gas detectors, with typically 1\% length changes causing potentially important effects over meter-scale detectors.  This swelling should be taken into consideration as part of the design of future experiments.

For structural elements, the effects of the gas absorption on mechanical strength may also be of interest. An interesting study in the future may be characterizing how the strength and durability of the materials changes with absorption, in order to inform stress analysis and mechanical deflection calculations.

Finally, we note that other materials are sometimes used in noble TPC experiments, including acrylic, G10 and Kapton. The significant swelling phenomena observed here, in addition to the non-trivial dependence of breakdown strength upon the gas environment, suggest that similar studies of these materials may be of value for future experiments.

\section*{Acknowledgements}
The NEXT Collaboration acknowledges support from the following agencies and institutions: the European Research Council (ERC) under the Advanced Grant 339787-NEXT; the European Union's Framework Programme for Research and Innovation Horizon 2020 (2014-2020) under the Marie SkÅodowska-Curie Grant Agreements No. 674896, 690575 and 740055; the Ministerio de Econom\'ia y Competitividad of Spain under grants FIS2014-53371-C04, the Severo Ochoa Program SEV-2014-0398 and the Mar\'ia de Maetzu Program MDM-2016-0692; the GVA of Spain under grants PROMETEO/2016/120 and SEJI/2017/011; the Portuguese FCT and FEDER through the program COMPETE, projects PTDC/FIS-NUC/2525/2014 and UID/FIS/04559/2013; and the U.S.\ Department of Energy under contracts number DE-AC02-07CH11359 (Fermi National Accelerator Laboratory), DE-FG02-13ER42020 (Texas A\&M), DE-SC0011686 (UTA) and DE-SC0017721 (UTA). F.~Psihas is supported by a ConTex postdoctoral fellowship. We also warmly acknowledge the Laboratorio Nazionale di Gran Sasso (LNGS) and the Dark Side collaboration for their help with TPB coating of various parts of the NEXT-White TPC. Finally, we are grateful to the Laboratorio Subterr\'aneo de Canfranc for hosting and supporting the NEXT experiment. 

\bibliographystyle{JHEP}
\bibliography{main}

\end{document}